\documentclass[11pt]{article}
\usepackage{axodraw}
\usepackage{epsfig}
\usepackage{amsfonts}
\usepackage{amsmath}
\usepackage{bbm}
\input pix.sty

\renewcommand{\eq}{eq.~}
\renewcommand{\eqs}{eqs.~}
\renewcommand{\se}{sec.~}

\renewcommand{\fig}{fig.~}
\renewcommand{\figs}{figs.~}

\newcommand{\tinymsbar}{{\overline{\mbox{\tiny\rm{MS}}}}}
\newcommand{\Lambdamsbar}{{\Lambda_\tinymsbar}}

\newcommand{\Nf}{N_{\rm f}}
\newcommand{\Nc}{N_{\rm c}}

\newcommand{\Tc}{T_{\rm c}}
\newcommand{\gB}{g_\rmii{B}}
\newcommand{\mE}{m_\rmii{E}}

\newcommand{\gammaE}{\gamma_\rmii{E}}

\newcommand{\rmO}{{\mathcal{O}}}
\newcommand{\bmu}{\bar\mu}
\newcommand{\CA}{\Nc}

\def\lsi{\raise0.3ex\hbox{$<$\kern-0.75em\raise-1.1ex\hbox{$\sim$}}}
\def\gsi{\raise0.3ex\hbox{$>$\kern-0.75em\raise-1.1ex\hbox{$\sim$}}}
\newcommand{\lsim}{\mathop{\lsi}}
\newcommand{\gsim}{\mathop{\gsi}}

\newcommand{\nF}{n_\rmii{F}}
\newcommand{\nB}{n_\rmii{B}}
 \renewcommand{\nF}[1]{n_\rmii{F{#1}}}
 \renewcommand{\nB}[1]{n_\rmii{B{#1}}}
\newcommand{\rmii}[1]{{\mbox{\tiny\rm{#1}}}}
\newcommand{\re}{\mathop{\mbox{Re}}}
\newcommand{\im}{\mathop{\mbox{Im}}}

\newcommand{\Tint}[1]{{\hbox{$\sum$}\!\!\!\!\!\!\!\int\,}_{\!\!\!\!\raise-0.9ex\hbox{$\scriptstyle{#1}$}}}
\newcommand{\Tinti}[1]{{{\Sigma}\!\!\!\!\raise0.3ex\hbox{$\int$}_\rmii{${#1}$}}}

\newcommand{\unit}{{\mathbbm{1}}} 
\newcommand{\bi}{\begin{itemize}}
\newcommand{\ei}{\end{itemize}}

\newcommand{\hide}[1]{ }

\def\TAsc(#1,#2)(#3,#4,#5)%
{\SetWidth{2.0}\CArc(#1,#2)(#3,#4,#5)\SetWidth{1.0}}
\def\Lwidth{3}

\def\TAgl(#1,#2)(#3,#4,#5){\SetWidth{2.0}\PhotonArc(#1,#2)(#3,#4,#5){\Lwidth}%
{6.283 #3 mul 360 div #4 #5 sub #4 #5 sub mul sqrt mul Tdensity mul}%
\SetWidth{1.0}}
\def\TLgl(#1,#2)(#3,#4){\SetWidth{2.0}\Photon(#1,#2)(#3,#4){\Lwidth}
{#1 #3 sub #1 #3 sub mul #2 #4 sub #2 #4 sub mul add sqrt Tdensity mul}%
\SetWidth{1.0}}

\renewcommand{\picc}[1]{\;\parbox[c]{60pt}{\begin{picture}(60,30)(0,0)
\SetWidth{1.0}\SetScale{1.3} #1 \end{picture}}\; }
\def\Lwidth{1.3}

\newcommand{\picu}[1]{\;\parbox[c]{30pt}{\begin{picture}(30,30)(0,0)
\SetWidth{1.0}\SetScale{1.0} #1 \end{picture}}\; }
\def\EleA{\picu{%
 \CArc(15,15)(15,0,360)%
 \Lgl(15,0)(15,30)%
 \COval(15,0)(2,2)(0){Black}{Black}%
 \COval(15,30)(2,2)(0){Black}{Black}%
}}
\def\EleB{\picu{%
 \CArc(15,15)(15,0,360)%
 \Lgl(15,0)(15,30)%
 \COval(15,0)(2,2)(0){Black}{Black}%
 \COval(15,30)(2,2)(0){Black}{Black}%
 \Agl(29,15)(8,100,260)%
}}
\def\EleBB{\picu{%
 \CArc(15,15)(15,0,360)%
 \Lgl(15,0)(15,30)%
 \Lgl(0,15)(12,15)%
 \Lgl(18,15)(30,15)%
 \COval(15,0)(2,2)(0){Black}{Black}%
 \COval(15,30)(2,2)(0){Black}{Black}%
}}
\def\EleC{\picu{%
 \CArc(15,15)(15,0,360)%
 \Agl(7,4)(8,-30,130)%
 \Agl(23,26)(8,150,310)%
 \COval(15,0)(2,2)(0){Black}{Black}%
 \COval(15,30)(2,2)(0){Black}{Black}%
}}
\def\EleD{\picu{%
 \CArc(15,15)(15,0,360)%
 \Agl(23,4)(8,50,210)%
 \Agl(23,26)(8,150,310)%
 \COval(15,0)(2,2)(0){Black}{Black}%
 \COval(15,30)(2,2)(0){Black}{Black}%
}}
\def\EleE{\picu{%
 \CArc(15,15)(15,0,360)%
 \Agl(40,30)(25,180,242)%
 \Agl(40,0)(25,118,137)%
 \Agl(40,0)(25,150,180)%
 \COval(15,0)(2,2)(0){Black}{Black}%
 \COval(15,30)(2,2)(0){Black}{Black}%
}}
\def\EleF{\picu{%
 \CArc(15,15)(15,0,360)%
 \Lgl(15,0)(15,30)%
 \Agl(40,40)(26,205,245)%
 \COval(15,0)(2,2)(0){Black}{Black}%
 \COval(15,30)(2,2)(0){Black}{Black}%
}}
\def\EleG{\picu{%
 \CArc(15,15)(15,0,360)%
 \Agl(40,15)(30,150,210)%
 \Agl(-10,15)(30,-30,30)%
 \COval(15,0)(2,2)(0){Black}{Black}%
 \COval(15,30)(2,2)(0){Black}{Black}%
}}
\def\EleH{\picu{%
 \CArc(15,15)(15,0,360)%
 \Lgl(15,0)(15,12)%
 \Agl(10,21.5)(10,-65,65)%
 \Agl(20,21.5)(10,115,245)%
 \COval(15,0)(2,2)(0){Black}{Black}%
 \COval(15,30)(2,2)(0){Black}{Black}%
}}
\def\EleI{\picu{%
 \CArc(15,15)(15,0,360)%
 \Lgl(15,0)(15,30)%
 \Lgl(15,15)(30,15)%
 \COval(15,0)(2,2)(0){Black}{Black}%
 \COval(15,30)(2,2)(0){Black}{Black}%
}}
\def\EleJ{\picu{%
 \CArc(15,15)(15,0,360)%
 \Lgl(15,0)(15,30)%
 \COval(15,0)(2,2)(0){Black}{Black}%
 \COval(15,30)(2,2)(0){Black}{Black}%
 \GCirc(15,15){4}{0.5}
}}
\newcommand{\pich}[1]{\!\!\parbox[c]{60pt}{\begin{picture}(60,60)(-30,0)
\SetWidth{1.0}\SetScale{0.7} #1 \end{picture}}\!\! }
\def\EleEa{\pich{%
 \CArc(0,30)(30,0,360)%
 \Lgl(0,0)(0,60)%
 \Lgl(0,30)(30,30)%
 \COval(0,0)(4,4)(0){Black}{Black}%
 \COval(0,60)(4,4)(0){Black}{Black}%
 \Text(0,-8)[c]{\bf $0$}%
 \Text(0,50)[c]{\bf $\tau$}%
 \Text(29,22)[c]{\bf $\tau'$}%
 \Text(-8,21)[c]{\bf $x$}%
}}
\def\EleEb{\pich{%
 \CArc(0,30)(30,0,360)%
 \Lgl(0,0)(0,60)%
 \Lgl(0,30)(-30,30)%
 \COval(0,0)(4,4)(0){Black}{Black}%
 \COval(0,60)(4,4)(0){Black}{Black}%
 \Text(0,-8)[c]{\bf $0$}%
 \Text(0,50)[c]{\bf $\tau$}%
 \Text(-29,22)[c]{\bf $\tau'$}%
 \Text(8,21)[c]{\bf $x$}%
}}
\newcommand{\piG}[1]{\;\parbox[c]{300pt}{\begin{picture}(150,40)(-175,-15)
\SetWidth{1.0}\SetScale{1.0} #1 \end{picture}}\;}
\def\LattGE{\piG{%
 \SetWidth{1.0} 
 \Line(-130,10)(-110,10)%
 \Line(-100,10)(-100,20)\Line(-100,20)(-90,20)%
 \Line(-85,10)(-75,10)\Line(-75,10)(-75,20)%
 \Line(-65,20)(-45,20)%
 \Line(-35,20)(-35,10)\Line(-35,10)(-25,10)%
 \Line(-20,20)(-10,20)\Line(-10,20)(-10,10)%
 \Vertex(-130,10){1}
 \Vertex(-110,10){1}
 \Vertex(-100,10){1}
 \Vertex(-90,20){1}
 \Vertex(-85,10){1}
 \Vertex(-75,20){1}
 \Vertex(-65,20){1}
 \Vertex(-45,20){1}
 \Vertex(-35,20){1}
 \Vertex(-25,10){1}
 \Vertex(-20,20){1}
 \Vertex(-10,10){1}
 \Text(-87.5,15)[c]{$-$}
 \Text(-22.5,15)[c]{$-$}
 \Text(-135,15)[c]{$\bigl\langle\,$}
 \Text(-105,15)[c]{$\bigl($}
 \Text(-70,15)[c]{$\bigr)$}
 \Text(-40,15)[c]{$\bigl($}
 \Text(-5,15)[c]{$\bigr)$}
 \Text(2,15)[c]{\small $+$}
 \Line(5,20)(25,20)%
 \Line(35,20)(35,10)\Line(35,10)(45,10)%
 \Line(50,20)(60,20)\Line(60,20)(60,10)%
 \Line(70,10)(90,10)%
 \Line(100,10)(100,20)\Line(100,20)(110,20)%
 \Line(115,10)(125,10)\Line(125,10)(125,20)%
 \Vertex(5,20){1}
 \Vertex(25,20){1}
 \Vertex(35,20){1}
 \Vertex(45,10){1}
 \Vertex(50,20){1}
 \Vertex(60,10){1}
 \Vertex(70,10){1}
 \Vertex(90,10){1}
 \Vertex(100,10){1}
 \Vertex(110,20){1}
 \Vertex(115,10){1}
 \Vertex(125,20){1}
 \Text(47.5,15)[c]{$-$}
 \Text(112.5,15)[c]{$-$}
 \Text(135,15)[c]{$\bigr\rangle$}
 \Text(30,15)[c]{$\bigl($}
 \Text(65,15)[c]{$\bigr)$}
 \Text(95,15)[c]{$\bigl($}
 \Text(130,15)[c]{$\bigr)$}
 \Text(-138,15)[r]{$\sum_{i=1}^{3}\re\tr$}
 \Line(-190,3)(140,3)
 \Text(-50,-10)[r]{$- 6 a^4 \re\tr\bigl\langle$}
 \Line(-48,-10)(15,-10)
 \Vertex(-48,-10){1}
 \Vertex(15,-10){1}
 \Text(20,-10)[c]{$\bigr\rangle$}
 \Text(-195,2)[r]{$G_\rmii{E}(\tau) = $}
 \LongArrow(175,-5)(175,15)%
 \LongArrow(175,-5)(155,-5)%
 \Text(157,1)[c]{\small $x_0$}
 \Text(167,13)[c]{\small $x_i$}
  }}
\newcommand{\piH}[1]{\;\parbox[c]{150pt}{\begin{picture}(0,40)(-175,-15)
\SetWidth{1.0}\SetScale{1.0} #1 \end{picture}}\;}
\def\LattC{\piH{%
 \SetWidth{1.0} 
 \Line(-130,10)(-90,10)%
 \Line(-90,10)(-90,20)%
 \Line(-90,20)(-50,20)%
 \Line(-50,20)(-50,10)%
 \Vertex(-130,10){1}
 \Vertex(-90,10){1}
 \Vertex(-90,20){1}
 \Vertex(-50,20){1}
 \Vertex(-50,10){1}
 \Text(-135,15)[c]{$\bigl\langle\,$}
 \Text(-45,15)[c]{$\bigr\rangle$}
 \Text(-140,15)[r]{$C(\tau) \equiv \re\tr$}
 \LongArrow(-110,0)(-90,0)%
 \LongArrow(-110,0)(-130,0)%
 \LongArrow(-70,0)(-50,0)%
 \LongArrow(-70,0)(-90,0)%
 \Text(-110,-10)[c]{\small $\beta-\tau$}
 \Text(-70,-10)[c]{\small $\tau$}
  }}
\makeatletter \@addtoreset{equation}{section} \makeatother
\renewcommand{\theequation}{\arabic{section}.\arabic{equation}}
\makeatletter
\renewcommand\section{\@startsection {section}{1}{\z@}%
                                   {-5.5ex \@plus -1ex \@minus -.2ex}
                                   {2.3ex \@plus.2ex}%
                                   {\normalfont\large\bfseries}}
\renewcommand\subsection{\@startsection{subsection}{2}{\z@}%
                                     {-3.25ex\@plus -1ex \@minus -.2ex}%
                                     {1.5ex \@plus .2ex}%
                                     {\normalfont\normalsize\bfseries}}
\renewcommand\thesection {\@arabic\c@section}
\renewcommand\thesubsection   {\thesection.\@arabic\c@subsection}
\renewcommand{\@seccntformat}[1]{%
\csname the#1\endcsname.\hspace{1.0em}}
\makeatother

\begin{document}

\flushbottom

\begin{titlepage}

\begin{flushright}
BI-TP 2010/16\\
\vspace*{1cm}
\end{flushright}
\begin{centering}
\vfill

{\Large{\bf
 Colour-electric spectral function at next-to-leading order
}} 

\vspace{0.8cm}

Y.~Burnier, 
M.~Laine, 
J.~Langelage, 
L.~Mether 

\vspace{0.8cm}
 

{\em
Faculty of Physics, University of Bielefeld, 
D-33501 Bielefeld, Germany\\}

\vspace*{0.8cm}

\mbox{\bf Abstract}
 
\end{centering}
 
\vspace*{0.3cm}
 
\noindent
The spectral function related to the correlator of two colour-electric 
fields along a Polyakov loop determines the momentum diffusion coefficient 
of a heavy quark near rest with respect to a heat bath. We compute this 
spectral function at next-to-leading order, $\rmO(\alpha_s^2)$, in the 
weak-coupling expansion. The high-frequency part of our result 
($\omega\gg T$), which is shown to be temperature-independent, is accurately 
determined thanks to asymptotic freedom; the low-frequency part of our 
result ($\omega\ll T$), in which Hard Thermal Loop resummation is needed in 
order to cure infrared divergences, agrees with a previously determined 
expression. Our result may help to calibrate the overall normalization 
of a lattice-extracted spectral function in a perturbative frequency 
domain $T \ll \omega \ll 1/a$, paving the way for a non-perturbative 
estimate of the momentum diffusion coefficient at $\omega\to 0$. 
We also evaluate the colour-electric Euclidean correlator, 
which could be directly compared with lattice simulations. 
As an aside we determine the Euclidean correlator in the lattice 
strong-coupling expansion, showing that through a limiting procedure 
it can in principle be defined also in the confined phase of pure 
Yang-Mills theory, even if a practical measurement could be very noisy there.  

\vfill

 
\vspace*{1cm}
  
\noindent
June 2010

\vfill

\end{titlepage}

%
\section{Introduction}

In order to theoretically model various
observations made at relativistic heavy ion collision 
experiments, it would be important to quantitatively 
determine a number of ``transport coefficients'' 
of Quantum Chromodynamics (QCD) at temperatures around a few
hundred MeV. Such transport coefficients include for instance
the shear and bulk viscosities and the thermal conductivity, 
which play a role in the hydrodynamic equations describing 
the expansion of the thermal fireball; as well as heavy flavour 
diffusion coefficients, the jet quenching parameter, and the electrical 
conductivity, which dictate how various ``probes'' 
propagate with respect to the hydrodynamically expanding thermal medium
(for recent reviews, see refs.~\cite{ga_rev,hbm_rev}). 

Among all the transport coefficients, the theoretically most accessible 
one appears to be the one related to the diffusive motion of heavy 
quarks (charm or bottom) near rest with respect to the thermal medium. 
In a classical Langevin picture, the heavy quarks satisfy 
the equation of motion ${\rm d}p_i/{\rm d}t = -\eta_\rmii{D} p_i + \xi_i$, 
where $p_i$ is their momentum and 
$\xi_i$ is a random force; the transport coefficient 
$\eta_\rmii{D}$ can be referred to as the ``kinetic thermalization rate''
or the ``drag coefficient''. The simplicity of its determination
is due to the fact that it can be fluctuation-dissipation 
related to another coefficient, 
$\kappa = \fr13 \int_{-\infty}^{\infty} \sum_i 
\langle \xi_i(t)\xi_i(0) \rangle$, called 
the ``momentum diffusion coefficient'' 
(the relation reads
$\eta_\rmii{D} = \kappa / (2 T M_\rmii{kin}) (1+\rmO(T/M_\rmii{kin}))$, 
where $M_\rmii{kin}$ is a certain heavy quark mass definition); 
$\xi_i$ can in turn be identified as the Lorentz force, 
which for small velocities is proportional to the electric field strength. 
Thereby $\kappa$ can be determined from the two-point correlator of 
colour-electric fields along a timelike Wilson line~\cite{ct,eucl}. 
The correlator of colour-electric fields is simpler 
to handle theoretically than that of the usual
bilinear quark currents, because fewer physical
scales enter the problem (heavy quark related scales 
like $M_\rmii{kin}$ do not appear) and because no transport peak is present. 

Within this framework, a number of results have 
recently been obtained. First of all, the leading perturbative
expression for $\kappa$~\cite{bt}--\cite{mt} has been supplemented by 
a next-to-leading order correction~\cite{chm}, which was furthermore
shown to be so large as to question the validity of the 
weak-coupling expansion. (This computation can be contrasted 
with the corresponding results for theories with gravity 
duals, in particular for strongly-coupled $\mathcal{N}=4$ Yang-Mills 
theory in the large-$\Nc$ limit~\cite{sea,ct}, in which case $\kappa$ appears 
to be much ``larger'' than according to leading-order QCD.) 
Second, the frequency dependence of the correlator 
$\kappa(\omega) = \fr13 \int_{-\infty}^{\infty} e^{i\omega t} \sum_i 
\langle \xi_i(t)\xi_i(0) \rangle$ has been 
determined within various frameworks, such as 
strongly coupled $\mathcal{N}=4$ Yang-Mills theory 
in the large-$\Nc$ limit~\cite{ct,ssg},
Hard Thermal Loop resummed perturbation theory~\cite{eucl}, 
as well as classical lattice gauge theory~\cite{mink}.
Gaining understanding on the frequency dependence is important
because a non-perturbative extraction of $\kappa$ from lattice QCD
would effectively necessitate an ansatz for the shape of $\kappa(\omega)$. 

Apart from these theoretical developments, heavy quark diffusion 
has also been embedded in phenomenological models
(see, e.g., refs.~\cite{mt,phen} and references therein). 
Comparing with experimental data~\cite{exp}, it appears
that the physical value of $\kappa$ would need to be 
significantly larger than the leading-order~\cite{bt}--\cite{mt}
or even the next-to-leading order one~\cite{chm}, perhaps 
in qualitative accord with the AdS/CFT result~\cite{sea,ct}
(cf.\ also ref.~\cite{ek} and references therein). 

The purpose of this paper is to address the frequency dependence
of $\kappa(\omega)$ within QCD and pure Yang-Mills theory, 
with an eventual lattice determination of
$\kappa \equiv \kappa(0)$ in mind. To this end, we determine
$\kappa(\omega)$ at next-to-leading order, $\rmO(\alpha_s^2)$, 
in the weak-coupling expansion. 
Thanks to asymptotic freedom, our result becomes increasingly accurate 
in the large-frequency limit where, as we show, the 
result is also independent of the temperature.
In fact, in several respects, the result at $\omega\gsim T$
is analogous to that
obtained a long time ago for the spectral function related
to the electromagnetic current~\cite{spectral}; nevertheless,
we are not aware of a previous computation of a spectral function 
at $\rmO(\alpha_s^2)$ consistently applicable in the whole frequency
range from $\omega\lsim \alpha_s^{1/2} T$ up to $\omega\gg T$.

The paper is organized as follows. 
In section~\ref{se:basic} we define the basic observable 
considered and introduce our notation. 
The main steps of the computation are outlined and 
to some extent also carried out in~section~\ref{se:outline}; 
a number of technical details are relegated to two appendices.  
Section~\ref{se:results} collects together various results, both 
in analytic form as well as numerically. Section~\ref{se:strong}
is complementary to the main body of this work, addressing
the colour-electric field correlator in the lattice 
strong-coupling expansion.
We summarize and offer an outlook in section~\ref{se:concl}.

%
\section{Basic definitions}
\la{se:basic}

The computation is carried out within continuum QCD, with 
$\Nc$ colours and $\Nf$ massless quark flavours. The theory 
is regularized dimensionally, and renormalized quantities 
are expressed in the $\msbar$ scheme. The only parameter
requiring renormalization is the gauge coupling; the bare gauge 
coupling is denoted by $\gB$, the renormalized one by $g$, 
and $\alpha_s \equiv g^2/4\pi$. The relation 
between $\gB$ and $g$ will only be needed at 1-loop order, 
\be
 \gB^2 = g^2 + \frac{g^4 \mu^{-2\epsilon}}{(4\pi)^2} 
 \frac{2\Nf - 11\Nc}{3\epsilon} + \rmO(g^6)
 \;, \la{gBare}
\ee
where $\mu$ is a scale parameter
related to ``minimal subtraction'', and the corresponding $\msbar$ scheme 
quantity is denoted by $\bmu^2 = 4 \pi e^{-\gammaE} \mu^2$.  The factor
$\mu^{-2\epsilon}$ is normally not displayed explicitly; the same holds
for corrections of $\rmO(\epsilon)$ which are not multiplied by 
divergent coefficients. A group theory factor appearing 
frequently is defined by $C_F \equiv (\Nc^2 - 1)/ 2 \Nc$.

The observable considered contains colour-electric 
fields and temporal Wilson lines. Employing sign 
conventions where the covariant derivative 
reads $D_\mu = \partial_\mu - i \gB A_\mu$, with $A_\mu$ 
a traceless and hermitean gauge field, we define a temporal 
Wilson line at spatial position $\vec{r}\equiv\vec{0}$ through
\be
 U(\tau_b,\tau_a) \equiv 
 \unit + i\gB 
 \int_{\tau_a}^{\tau_b}\! {\rm d}\tau\, A_0(\tau,\vec{0}) 
 + (i\gB)^2 \int_{\tau_a}^{\tau_b} \! {\rm d}\tau \, 
 \int_{\tau_a}^\tau \! {\rm d}\tau' \, 
 A_0(\tau,\vec{0}) A_0(\tau',\vec{0}) + \ldots 
 \;. \la{Pdef}
\ee
A colour-electric field strength is defined through
\be
 \gB E_i \equiv i [D_0, D_i] 
 \;. 
\ee
With this notation, the Euclidean correlation function considered~\cite{eucl}
can be expressed as 
\be
 G_\rmii{E}(\tau) \equiv - \fr13 \sum_{i=1}^{3-2\epsilon} 
 \frac{
  \Bigl\langle
   \re\tr \Bigl[
      U(\beta,\tau) \, 
      \gB E_i(\tau,\vec{0}) \, U(\tau,0) \, 
      \gB E_i(0,\vec{0})
   \Bigr] 
  \Bigr\rangle
 }{
 \Bigl\langle
   \re\tr [U(\beta,0)] 
 \Bigr\rangle
 }
 \;, \la{GE_def}
\ee
where $\beta\equiv 1/T$ is the inverse temperature. 
Note that we keep a fixed ``3'' in the denominator even 
in the presence of dimensional regularization; this has no effect 
on the final renormalized result but does affect the finite parts
of divergent intermediate expressions. 

The spectral function corresponding to 
$G_\rmii{E}(\tau)$ can be determined  from 
\ba
  \tilde G_\rmii{E}(\omega_n) & \equiv & 
  \int_0^\beta \! {\rm d}\tau \, e^{i \omega_n\tau } G_\rmii{E}(\tau)
  \;, \la{tildeGE} \\ 
  \rho_\rmii{E}(\omega) & = & 
  \im \tilde G_\rmii{E}(\omega_n \to -i [\omega + i 0^+])
  \;,
 \la{disc}
\ea
where the analytic continuation becomes unique by requiring
powerlike (not exponential) growth at asymptotic frequencies;  
the momentum diffusion coefficient then follows from 
\be
 \kappa \equiv \lim_{\omega\to 0} \frac{2 T \rho_\rmii{E}(\omega)}{\omega} 
 \;. \la{kappa_def_3}
\ee
If only a numerical determination of $G_\rmii{E}(\tau)$ is available, 
the task would be to invert the relation
\be
 G_\rmii{E}(\tau) = 
 \int_0^\infty
 \frac{{\rm d}\omega}{\pi} \rho_\rmii{E}(\omega)
 \frac{\cosh \left(\frac{\beta}{2} - \tau\right)\omega}
 {\sinh\frac{\beta \omega}{2}} 
 \;, \la{int_rel} 
\ee
for which various practical recipes have been proposed; 
see, e.g., refs.~\cite{recent}. (Probably the asymptotic
large-$\omega$ behaviour of $\rho_\rmii{E}(\omega)$
is needed as input for the inversion.)

%
\begin{figure}[t]
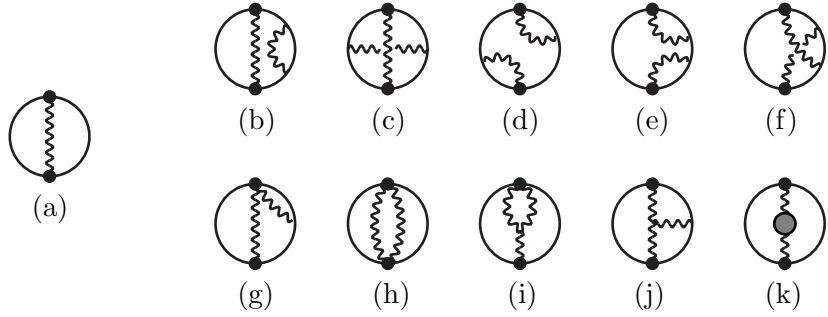


\hspace*{1.5cm}%
\begin{minipage}[c]{3cm}
\begin{eqnarray*}
&& 
 \hspace*{-1cm}
 \EleA 
\\[1mm] 
&& 
 \hspace*{-0.6cm}
 \mbox{(a)} 
\end{eqnarray*}
\end{minipage}%
\begin{minipage}[c]{10cm}
\begin{eqnarray*}
&& 
 \hspace*{-1cm}
 \EleB \quad\; 
 \EleBB \quad\; 
 \EleC \quad\; 
 \EleD \quad\; 
 \EleE \quad\; 
\\[1mm] 
&& 
 \hspace*{-0.6cm}
 \mbox{(b)} \hspace*{1.26cm}
 \mbox{(c)} \hspace*{1.26cm}
 \mbox{(d)} \hspace*{1.26cm}
 \mbox{(e)} \hspace*{1.26cm}
 \mbox{(f)} 
\\[5mm] 
&& 
 \hspace*{-1cm}
 \EleF \quad\; 
 \EleG \quad\; 
 \EleH \quad\; 
 \EleI \quad\; 
 \EleJ \quad 
\\[1mm] 
&& 
 \hspace*{-0.6cm}
 \mbox{(g)} \hspace*{1.3cm}
 \mbox{(h)} \hspace*{1.28cm}
 \mbox{(i)} \hspace*{1.28cm}
 \mbox{(j)} \hspace*{1.32cm}
 \mbox{(k)} 
\end{eqnarray*}
\end{minipage}

\caption[a]{\small 
The graphs contributing to the colour-electric field correlator, 
$G_\rmii{E}(\tau)$ defined in \eq\nr{GE_def}, up to $\rmO(g^4)$. 
The big circle denotes a Wilson line wrapping around the time 
direction; the small dots the colour-electric field strengths; 
and the grey blob the 1-loop gauge field self-energy. Graphs obtained 
with trivial ``reflections'' from those shown have been 
omitted from the figure.} 
\la{fig:Poly}
\end{figure}
%

The graphs contributing to the numerator of $G_\rmii{E}$
are shown up to $\rmO(g^4)$ 
in \fig\ref{fig:Poly}. As is typically 
the case in finite-temperature QCD, however, certain subclasses 
of higher-order graphs may need to be considered as well. 
This can be achieved with the so-called Hard Thermal Loop (HTL)  
effective theory~\cite{htl}; where necessary, the observables 
will then be expressed as 
\be
 G_\rmii{E} = \Bigl[ 
          (G_\rmii{E})_\rmii{QCD} - 
          (G_\rmii{E})_\rmii{HTL} \Bigr]_\rmii{naive} + 
          \Bigl[ (G_\rmii{E})_\rmii{HTL} \Bigr]_\rmii{resum} 
 \;, \la{resum}
\ee 
where ``naive'' refers to an unresummed computation
and ``resum'' to a resummed one. 
The unresummed difference in the first square brackets is infrared 
finite (provided that the correct low-energy effective description 
is used), and can be computed in naive perturbation theory, 
i.e.\ from the graphs of \fig\ref{fig:Poly}; in an effective
field theory language, it corresponds to an ultraviolet matching
coefficient. The last term of \eq\nr{resum} is 
infrared sensitive, but can be addressed within the simplified 
effective description rather than with full QCD.

%
\section{Outline of the computation}
\la{se:outline}

%
\subsection{Leading order analysis}

In order to illustrate the steps of the computation, we start 
by explicitly deriving the (well-known) leading order result 
for the spectral function, corresponding to the contribution 
of graph (a) of \fig\ref{fig:Poly}. A direct evaluation of 
the Wick contractions leads to 
\be
 G_\rmii{E}^{(2)}(\tau) = 
 \frac{\gB^2C_F}{3} \sum_i 
 \Bigl\{
   \partial_0^2 \Delta_{ii}(\tau,\vec{0})+ 
   \partial_i^2 \Delta_{00}(\tau,\vec{0})- 
 \partial_0 \partial_i \Bigl[ 
   \Delta_{0i}(\tau,\vec{0}) + \Delta_{i0}(\tau,\vec{0})
 \Bigr]
 \Bigr\}
 \;, 
\ee
where $\Delta_{\mu\nu}$ is the gauge field propagator in configuration space.
Expressing the propagator in momentum space, 
\be
 \Delta_{\mu\nu}(\tau,\vec{x}) = 
 \Tint{K} e^{i k_n \tau + i \vec{k}\cdot\vec{x}}
 \tilde\Delta_{\mu\nu}(K)
 \;, 
\ee
where $\Tinti{K} \equiv T \sum_{k_n} \int_{\vec{k}}$ and 
$k_n$ is a bosonic Matsubara frequency, $k_n = 2\pi T n$, $n\in\mathbbm{Z}$; 
and inserting the propagator of the general covariant gauge, 
\be
 \tilde\Delta_{\mu\nu}(K) = \frac{\delta_{\mu\nu}}{K^2} + 
 \frac{\tilde \xi  K_\mu K_\nu}{(K^2)^2}
 \;, \la{G_xi}
\ee
the longitudinal part proportional to the gauge parameter, $\tilde\xi$,
is seen to drop out. The remaining expression reads 
\be
  G_\rmii{E}^{(2)}(\tau) = 
 - \frac{\gB^2C_F} {3}  \Tint{K} e^{i k_n\tau}
 \frac{(D-1) k_n^2 + \vec{k}^2}{K^2}
 \;,
\ee
where $D=4-2\epsilon$ is the spacetime dimensionality. 
Writing $k_n^2 = K^2 - \vec{k}^2$, omitting the scaleless integral
$\int_{\vec{k}} 1$ due to properties of dimensional regularization, 
and making use of 
\be
 T\sum_{k_n} \frac{e^{i k_n\tau}}{k_n^2 + k^2}
 = \frac{\nB{}(k)}{2k}\Bigl[e^{(\beta-\tau)k} + e^{\tau k} \Bigr]
 \;, \quad 0 \le \tau \le \beta
 \;, \la{sum}
\ee
where $k\equiv |\vec{k}|$, we get 
\be
 G_\rmii{E}^{(2)}(\tau) = 
 \frac{\gB^2 C_F (D-2)}{6}
 \int_{\vec{k}} k\, \nB{}(k) \Bigl[e^{(\beta-\tau)k}+e^{\tau k} \Bigr]
 \;. 
\ee
The Fourier transform as in \eq\nr{tildeGE} yields 
\ba
 \tilde G_\rmii{E}^{(2)}(\omega_n) & = & 
 \frac{\gB^2 C_F (D-2)}{6}
 \int_{\vec{k}} k 
 \biggl[ 
   \frac{1}{k-i\omega_n} + \frac{1}{k+i\omega_n}
 \biggr]
 \;, 
\ea
and taking a subsequent discontinuity according to \eq\nr{disc} we obtain 
\ba
 \rho_\rmii{E}^{(2)}(\omega) & = & 
 \frac{\gB^2 C_F (D-2) \pi}{6}
 \int_{\vec{k}} k 
 \Bigl[ 
   \delta(k-\omega) - \delta(k+\omega) 
 \Bigr]
 \;.  \la{rhoE2_bare_1}
\ea
The final integral over $\vec{k}$ can now be trivially carried out. 
For $D=4$ the result reads
\be
 \rho_\rmii{E}^{(2)}(\omega) = \frac{g^2 C_F}{6\pi} \omega^3
 \;, \la{rhoE2}
\ee
where we also expanded the bare gauge coupling according 
to \eq\nr{gBare}.

%
\subsection{Wick contractions at next-to-leading order}

Proceeding to the level $\rmO(g^4)$, 
the structures in the graphs (b)--(k) of \fig\ref{fig:Poly}
need to be considered. (There is also a disconnected contribution 
originating from the numerator of \eq\nr{GE_def}, as will be 
specified presently.) Given that the observable is gauge invariant, 
the full result must be independent of the gauge parameter $\tilde\xi$; 
nevertheless the individual diagrams contain non-trivial
terms proportional to $\tilde\xi$ and $\tilde\xi^2$. We have checked
explicitly that both structures cancel in the sum, and display 
intermediate expressions only for the Feynman gauge, $\tilde\xi = 0$. 
(As a further cross-check, the small-frequency limit will be 
compared against that of an independent computation in the Coulomb gauge,
cf.\ \se\ref{ss:ir}).

Considering first the structures (b)--(f) of \fig\ref{fig:Poly}, 
and combining them with the contribution of the denominator of \eq\nr{GE_def},
{\it viz.}\ 
\be
 \Bigl\langle
   \re\tr [U(\beta,0)] 
 \Bigr\rangle
 = \Nc 
 \Bigl[ 
   1 - \gB^2 C_F \int_0^\beta \! {\rm d}\tau \! \int_0^\tau \! {\rm d}\tau' 
   \, \Delta_{00}(\tau',\vec{0})
 \Bigr]  + \rmO(\gB^4)
 \;, \la{norm}
\ee
we obtain
\ba
 \delta_\rmii{(b-f)} G_\rmii{E}^{(4)}(\tau)
 & = & 
 \frac{\gB^4 C_F \CA}{6}
 \Bigl[(D-1) \partial_\tau^2 + \nabla^2 \Bigr] G(\tau,\vec{0})
 \int_{\tau}^{\beta} \! {\rm d}\tau' \int_0^\tau \! {\rm d}\tau''
 G(\tau'-\tau'',\vec{0})
 \nn & = &  
 \frac{\gB^4 C_F \CA}{3}
 \Tint{K} \frac{e^{i k_n\tau}}{K^2}
 \Bigl[-(D-1) K^2 + (D-2) k^2 \Bigr]
 \Tint{Q'} \frac{e^{i q_n\tau} - 1}{q_n^2 Q^2} 
 \;. \la{dGEbf}
\ea
Here we denoted 
$G(x_0,\vec{x})\equiv [\Delta_{00}(x_0,\vec{x})]_{\tilde\xi = 0}$, i.e.\ 
\be
 G(x_0,\vec{x}) =
 \Tint{K} \frac{e^{i k_n x_0 + i \vec{k}\cdot\vec{x}}}{k_n^2 + \vec{k}^2}
 \;, 
 \la{Gx}
\ee
and moved to momentum space, making use of 
\ba
 \int_0^\tau \! {\rm d}\tau' \, 
 e^{i q_n \tau'}
 & = & 
  \delta_{q_n} \, \tau  
 + 
 (1-\delta_{q_n})
 \frac{1}{i q_n} \Bigl( e^{i q_n \tau} - 1 \Bigr)
 \;, \la{tri1} 
\ea
etc, where $q_n$ is a bosonic Matsubara frequency 
and $\delta_{q_n} \equiv \delta_{n,0}$ is a (periodic) 
Kronecker delta function. 
It can be verified that the Matsubara zero-mode 
part of \eq\nr{tri1} does not contribute in dimensional 
regularization in \eq\nr{dGEbf}; the prime 
in $\Tinti{Q'}$ indicates that it is omitted. 
The term with $(D-1)K^2$ in  \eq\nr{dGEbf} does
not contribute either, because the scaleless integral
$\int_{\vec{k}} 1$ vanishes in dimensional regularization.

The contributions of the graphs (g) and (h) of \fig\ref{fig:Poly}
can be worked out in a similar way. A straightforward computation yields
\ba
 \delta_\rmii{(g)} G_\rmii{E}^{(4)}(\tau)
 & = & 
 \frac{\gB^4 C_F \CA (D-1)}{3} \partial_\tau G(\tau,\vec{0})
 \int_{\tau}^{\beta-\tau} \! {\rm d}\tau' \, G(\tau',\vec{0})
 \nn & = & 
 -  \frac{2 \gB^4 C_F \CA (D-1)}{3}
 \Tint{K} \frac{i k_n e^{i k_n\tau}}{K^2}
 \Tint{Q'}\frac{e^{i q_n\tau} - 1}{i q_n Q^2}
 \;, \la{dGEg} \\ 
 \delta_\rmii{(h)} G_\rmii{E}^{(4)}(\tau)
 & = & 
 - \frac{\gB^4 C_F \CA (D-1)}{3} \Bigl[ G(\tau,\vec{0}) \Bigr]^2 
 \nn & = & 
  -  \frac{\gB^4 C_F \CA (D-1)}{3}
 \Tint{K,Q} \frac{e^{i k_n\tau}}{Q^2(K-Q)^2}
 \;. \la{dGEh}
\ea

The graphs (i) and particularly (j) of \fig\ref{fig:Poly} are 
somewhat more complicated than the ones discussed so far. 
For (i) a few steps lead to the momentum space expression
\be
 \delta_\rmii{(i)} G_\rmii{E}^{(4)}(\tau) = 
 \gB^4C_F \CA  \Tint{K} e^{i k_n\tau}
 \Tint{Q} 
   \biggl[  \frac{D-1}{Q^2(K-Q)^2}
           - \frac{(D-2)k^2}{K^2 Q^2(K-Q)^2}
   \biggr]
 \;, \la{dGEi}
\ee
while for (j) it is perhaps simplest to remain in configuration space: 
\be
 \delta_\rmii{(j)} G_\rmii{E}^{(4)}(\tau) =
 -   \delta_\rmii{(g)} G_\rmii{E}^{(4)}(\tau)
 + \frac{\gB^4C_F \CA}{6} \, \mathcal{I}_5(\tau)
 \;. \la{dGEj}
\ee
Here we carried out a partial integration and made use of 
$
 (\partial_0^2 + \nabla^2)G(x_0,\vec{x}) = -\delta^{(4)}(x) 
$, 
in order to identify
a term that cancels against that in \eq\nr{dGEg}; 
and expressed the remainder through the integral 
\ba
 && \hspace*{-1cm} \mathcal{I}_5(\tau)  \equiv 
 \int_x \biggl[\int_{\tau}^{\beta} \! {\rm d}\tau'  - 
 \int_{0}^{\tau} \! {\rm d}\tau' \biggr] 
 \biggl\{ 
  \nn 
  &&  \!\!
  \partial_i G(x_0-\tau',\vec{x}) 
   \Bigl[ 
          \partial_0 G(x_0-\tau,\vec{x}) \partial_i G(x_0,\vec{x}) - 
          \partial_i G(x_0-\tau,\vec{x}) \partial_0 G(x_0,\vec{x}) \Bigr]
  \nn & + & \!\! (D-2) G(x_0-\tau',\vec{x}) 
  \Bigl[ 
          \partial_0 G(x_0-\tau,\vec{x}) \nabla^2 G(x_0,\vec{x}) - 
          \nabla^2 G(x_0-\tau,\vec{x}) \partial_0 G(x_0,\vec{x}) \Bigr]
 \biggr\}
 \;. \nn \la{I5_1} \la{I5_def}
\ea
The subscript ``5'' refers to the fact that this is the most complicated
among the 5 independent structures that appear in our result
(the other ones are defined in \eqs\nr{I1_def}--\nr{I4_def}).\footnote{%
 In fact $\mathcal{I}_5$ could be subdivided into further independent
 structures, but here we keep it as one entity. 
 } 
A corresponding momentum space expression, albeit with no care 
taken of the Matsubara zero mode contribution, 
is given in \eq\nr{I5_3} of appendix~A.

Inspecting finally the self-energy contribution, 
graph (k) in \fig\ref{fig:Poly}, we recall that 
making use of the substitution $Q\to K-Q$ in order to simplify 
the numerator, the Feynman gauge self-energy has the form 
\ba
 \Pi_{\mu\nu}(K) & = & 
 \frac{\gB^2 \CA}{2} 
 \Tint{Q} 
 \frac{ 
   \delta_{\mu\nu}\Bigl[ -4 K^2 + 2 (D-2) Q^2 \Bigr]
   + (D+2) K_\mu K_\nu - 4 (D-2) Q_\mu Q_\nu
 }{Q^2(K-Q)^2}
   \nn &  - &  
 \gB^2 \Nf\, \Tint{\{ Q \} }
 \frac{ 
   \delta_{\mu\nu} \Bigl[ -K^2 + 2 Q^2 \Bigr]
   + 2 K_\mu K_\nu - 4 Q_\mu Q_\nu  
 }{Q^2(K-Q)^2}
 \;, \la{Pi_xi}
\ea
and the corresponding contribution to $G_\rmii{E}(\tau)$ reads
\be
 \delta_\rmii{(k)} G_\rmii{E}^{(4)}(\tau)
 = \frac{\gB^2C_F}{3}  \Tint{K} 
   \frac{e^{i k_n\tau}}{(K^2)^2} 
   {\sum_i}
   (k_n \delta_{i\mu} - k_i \delta_{0\mu})
   (k_n \delta_{i\nu} - k_i \delta_{0\nu}) \Pi_{\mu\nu}(K)
  \;. \la{selfE}
\ee
Following the usual convention,  $\Tinti{ \{ Q \} }$ 
in \eq\nr{Pi_xi} means that the Matsubara
frequency is fermionic, i.e.\ $\{ q_n\} =  (2 n +1) \pi T $, 
with $n \in \mathbbm{Z}$; 
also, in \eq\nr{selfE}, $K=(k_n,k_i)$.

Combining \eqs\nr{Pi_xi} and \nr{selfE}, 
the transverse projectors eliminate the parts proportional 
to $K_\mu K_\nu$ from $\Pi_{\mu\nu}$, while the parts 
proportional to $Q_\mu Q_\nu$ can be expressed as  
\be
  {\sum_i}
   (k_n \delta_{i\mu} - k_i \delta_{0\mu})
   (k_n \delta_{i\nu} - k_i \delta_{0\nu}) Q_\mu Q_\nu 
  = (K-Q)^2 k_n q_n + Q^2 k_n (k_n-q_n) + K^2 q_n (q_n - k_n)
 \;. 
\ee 
The first two terms are odd in the summation variable $q_n$
or $k_n-q_n$, respectively, and give no contribution in \eq\nr{selfE},
so that only the last term matters. 
In addition there is a contribution from the term proportional 
to $\delta_{\mu\nu}$ in $\Pi_{\mu\nu}$ which we write as 
\be
  {\sum_i}
   (k_n \delta_{i\mu} - k_i \delta_{0\mu})
   (k_n \delta_{i\nu} - k_i \delta_{0\nu}) \delta_{\mu\nu}
  = (D-1) K^2 - (D-2) k^2
 \;. 
\ee
In total, then, 
\ba
 \delta_\rmii{(k)} G_\rmii{E}^{(4)}(\tau)
 & =&  \frac{\gB^4C_F}{3}  \Tint{K} e^{i k_n\tau}
 \nn & & \; \times
 \biggl\{
  \frac{\CA}{2}
  \Tint{Q} 
   \biggl[ - \frac{4(D-1)}{Q^2(K-Q)^2}
           + \frac{4(D-2)k^2}{K^2 Q^2(K-Q)^2}
           - \frac{4(D-2) q_n(q_n-k_n)}{K^2 Q^2(K-Q)^2}
  \nn & & \hspace*{2.0cm}
           + \, \frac{2(D-2)}{Q^2} \biggl( \frac{D-1}{K^2} 
                            - \frac{(D-2) k^2}{(K^2)^2} \biggr)
   \biggr]
  \nn & & \;
  -   \Nf\, 
  \Tint{\{ Q \} }
   \biggl[ 
           -  \frac{(D-1)}{Q^2(K-Q)^2}
           +  \frac{(D-2)k^2}{K^2 Q^2(K-Q)^2}
           - \frac{4q_n(q_n-k_n)}{K^2 Q^2(K-Q)^2}
  \nn & & \hspace*{2.0cm}
           + \, \frac{2}{Q^2} \biggl( \frac{D-1}{K^2} 
                        - \frac{(D-2) k^2}{(K^2)^2} \biggr)
   \biggr]
 \biggr\}
 \;. \la{dGEk}
\ea

We now sum together 
\eqs\nr{dGEbf}, \nr{dGEg}--\nr{dGEj}, \nr{dGEk}.
It can be noted that, in dimensional regularization, 
\be
  \int_{\vec{k}} \frac{k^2}{(k_n^2 + k^2)^2} = 
 \frac{D-1}{2}
  \int_{\vec{k}} \frac{1}{k_n^2 + k^2} 
 \;.  
\ee 
As a consequence the factorized structures on the 3rd and 5th 
rows of \eq\nr{dGEk} are proportional to $4-D = 2\epsilon$, 
and since the sum-integrals
$\Tinti{Q}1/Q^2 = T^2/12 + \rmO(\epsilon)$ and
$\Tinti{ \{Q\} }1/Q^2 = -T^2/24 + \rmO(\epsilon)$
are finite in dimensional regularization, there is no contribution
from these terms. Defining furthermore 
\ba
 \mathcal{I}_1(\tau) & \equiv & \Tint{K,Q} \frac{e^{i k_n\tau}}{Q^2(K-Q)^2}
 \;, \la{I1_def}  \\
 \mathcal{I}_2(\tau) & \equiv & \Tint{K,Q} 
 \frac{k^2 e^{i k_n\tau}}{K^2 Q^2(K-Q)^2}
 \;, \la{I2_def}  \\
 \mathcal{I}_3(\tau) & \equiv & \Tint{K,Q} 
 \frac{q_n(q_n-k_n) e^{i k_n\tau}}{K^2 Q^2(K-Q)^2}
 \;, \la{I3_def}  \\
 \mathcal{I}_4(\tau) & \equiv & 
 \Tint{K} \frac{k^2 e^{i k_n\tau}}{K^2}
 \Tint{Q'} \frac{e^{i q_n\tau} - 1}{q_n^2 Q^2} 
 \;, \la{I4_def}
\ea 
and correspondingly 
$\mathcal{I}_{\{1\}}$, 
$\mathcal{I}_{\{2\}}$, 
$\mathcal{I}_{\{3\}}$ 
for the cases that $Q$ is fermionic, and recalling 
$\mathcal{I}_5$ from \eq\nr{I5_1}, we can write 
the complete (unresummed) contribution as 
\ba
 \Bigl[ \delta G_\rmii{E}^{(4)}(\tau) \Bigr]_\rmii{QCD, naive} & = & 
 \frac{\gB^4 C_F \CA}{3}
 \biggl\{
  (D-2) \Bigl[ -\mathcal{I}_2(\tau) 
 - 2 \mathcal{I}_3(\tau) + \mathcal{I}_4(\tau) \Bigr]
  + \frac{1}{2} \mathcal{I}_5(\tau) 
 \biggr\} 
 \nn & + & 
 \frac{\gB^4 C_F \Nf}{3}
 \biggl[  
   (D-1) \mathcal{I}_{\{1\}}(\tau)
   - (D-2) \mathcal{I}_{\{2\}}(\tau)
   + 4 \mathcal{I}_{\{3\}}(\tau)
 \biggr]
 \;. \la{dGE4}
\ea

%
\subsection{Matsubara sums, spatial integrals, spectral function}

The next steps are to carry out the Matsubara sums over 
$k_n,q_n$ (generalizing \eq\nr{sum}); to Fourier transform 
with respect to $\tau$ (\eq\nr{tildeGE}); to take 
the discontinuity across the real axis (\eq\nr{disc}); 
and to carry out the remaining spatial integrals. 
We illustrate these steps for one of the 
structures appearing, $\mathcal{I}_{2}$ of \eq\nr{I2_def}, 
in some detail 
in appendices~\ref{ss:mat}--\ref{ss:fz}.  
An alternative derivation of the ``vacuum part'' 
of the most complicated structure, $\mathcal{I}_{5}$, 
can be found in appendix~\ref{ss:alt}. The final 
results are collected in appendix~\ref{ss:res}.
Inserting the expressions from \eqs\nr{finalnewI1}--\nr{finalI5}
of appendix~\ref{ss:res}, 
together with $D=4-2\epsilon$, 
we thus get the unresummed bare contribution to the 
colour-electric spectral function:
\ba
 \Bigl[ \delta \rho_\rmii{E}^{(4)}(\omega) \Bigr]_\rmii{QCD, naive} & = & 
 \frac{\gB^4 C_F \Nc}{3}
 \biggl\{
    2(1-\epsilon) \Bigl[ 
             -\tilde{\mathcal{I}}_2(\omega) 
             -2\tilde{\mathcal{I}}_3(\omega)
             +\tilde{\mathcal{I}}_4(\omega) 
          \Bigr] 
    + \fr12 \tilde{\mathcal{I}}_5(\omega) 
 \biggr\}
 \nn 
 & + & 
 \frac{\gB^4 C_F \Nf}{3}
 \biggl[  
   (3-2\epsilon) \tilde{\mathcal{I}}_{\{1\}}(\omega)
   - 2(1-\epsilon) \tilde{\mathcal{I}}_{\{2\}}(\omega)
   + 4 \tilde{\mathcal{I}}_{\{3\}}(\omega)
 \biggr]
 \la{rhoEmaster}
\ea
\ba
 & = & 
 \frac{\gB^4 C_F \Nc}{3} \biggl\{ 
 \frac{\omega^3 \mu^{-4\epsilon}}{(4\pi)^3}
 \biggl[ \frac{22}{3} \biggl( 
  \frac{1}{\epsilon} + 2 \ln\frac{\bmu^2}{4\omega^2} \biggr)
 + \frac{364}{9} - \frac{16\pi^2}{3}
 \biggr]
 \nn & & \; 
 + \frac{1}{4\pi^3} \int_0^\infty \! {\rm d} q \, \nB{}(q) 
 \biggl[
   (q^2 +2 \omega^2) \ln \left| \frac{q+w}{q-w} \right| 
   + q \omega \biggl( \ln\frac{|q^2 - \omega^2|}{\omega^2} - 1 \biggr)
  \nn & & \hspace*{3cm} +\;
      \frac{\omega^4}{q} 
     \mathbbm{P} \biggl( \frac{1}{q+\omega} \ln \frac{q + \omega}{\omega} 
     +   \frac{1}{q-\omega} \ln \frac{\omega}{|q - \omega|} \biggr) \biggr]
 \biggr\}  \hspace*{0.7cm}
 \nn & + &   
  \frac{\gB^4 C_F \Nf}{3} \biggl\{ 
 \frac{\omega^3 \mu^{-4\epsilon}}{(4\pi)^3}
 \biggl[ -\frac{4}{3} \biggl( 
  \frac{1}{\epsilon} + 2 \ln\frac{\bmu^2}{4\omega^2} \biggr)
 - \frac{52}{9} 
 \biggr]
 \nn & & \; 
 + \frac{1}{4\pi^3} \int_0^\infty \! {\rm d} q \, \nF{}(q) 
 \biggl[
   \Bigl(q^2 +\frac{\omega^2}{2} \Bigr) \ln \left| \frac{q+w}{q-w} \right| 
   + q \omega \biggl( \ln\frac{|q^2 - \omega^2|}{\omega^2} - 1 \biggr)
 \biggr\} \;. \hspace*{0.7cm}
 \la{rhoEbare}
\ea
To renormalize this, we need the generalization of \eq\nr{rhoE2}
with effects of $\rmO(\epsilon)$ included, which can be obtained from
\eq\nr{fz_basic}, taking into account the $D-2$ from 
\eq\nr{rhoE2_bare_1}: 
\be
 \Bigl[ \rho_\rmii{E}^{(2)}(\omega) \Bigr]_\rmii{QCD, naive} =  
 \frac{\gB^2 C_F }{6\pi} \; 
 \omega^3 \mu^{-2\epsilon} \biggl[
  1 +  \epsilon \biggl(\ln\frac{\bmu^2}{4\omega^2} + 1 \biggr)  
 \biggr]
 \;. 
 \la{rhoE2_bare}
\ee
Inserting here the bare gauge coupling from \eq\nr{gBare} 
and re-expanding in terms of the renormalized gauge coupling, we get 
an additional contribution which removes 
the divergences from \eq\nr{rhoEbare}:
\ba
 \Bigl[ \delta \rho_\rmii{E}^{(4,\rmii{ct})}(\omega)
 \Bigr]_\rmii{QCD, naive} & = & 
 \frac{g^4 C_F }{3}  
 \frac{\omega^3 \mu^{-4\epsilon}}{(4\pi)^3}
 \biggl( - \fr{22\Nc}3 + \fr{4\Nf}3 \biggr) \biggl( \frac{1}{\epsilon} + 
 \ln\frac{\bmu^2}{4\omega^2} + 1  \biggr) 
 \;. \la{ct}
\ea
The sum of \eqs\nr{rhoEbare} and \nr{ct} is our full result 
for the unresummed next-to-leading order contribution 
to $\rho_\rmii{E}(\omega)$; it is reproduced 
in complete form in \eq\nr{rho_hard} below.

%
\subsection{Hard Thermal Loop resummation}
\la{ss:HTL}

The result obtained so far (given explicitly in \eq\nr{rho_hard} below)
contains no infrared divergences, so that in the regime 
$\omega\gsim T$ there is no need for 
a resummation to be carried out. Nevertheless, we would like to 
extend the applicability of the result also to the regime 
$\omega\lsim gT$ in which collective phenomena like Debye screening
play an important role. The reason for this desire 
is on one hand formal, allowing us 
to cross-check that the framework is in principle well-defined 
in this limit as well; and on the other hand practical, permitting us 
to compare with existing results in the literature. In the regime
$\omega\lsim gT$, logarithmic divergences do appear and a resummation
becomes necessary. 

The way to consistently carry out the resummation is dictated 
by \eq\nr{resum}. The practical computations are somewhat cumbersome; 
given that they also have no bearing on our main result (i.e.\ the behaviour
of the spectral function in the regime $\omega\gsim T$), we relegate
the discussion to appendix~B. The main results can be found in 
\eqs\nr{rho_soft}, \nr{wdep} below, with a notation as introduced 
in \eq\nr{rho_full}.

%
\section{Results}
\la{se:results}

In this section we collect together the results from the previous section
and the appendices,  
and write down our final formulae for 
the colour-electric spectral function.

%
\subsection{Full expression}
\la{ss:full}

In order to write down the full result, we rephrase \eq\nr{resum} as
\ba
 \rho_\rmii{E}(\omega) & = &
 \Bigl[ 
 \Bigl( \rho_\rmii{E} \Bigr)_\rmii{QCD}
 -      \Bigl( \rho_\rmii{E} \Bigr)_\rmii{HTL}
 \Bigr]_\rmii{naive}
 + \Bigl( \rho_\rmii{E} \Bigl)_\rmii{HTL, resum} 
 \nn 
 & = &  
 \Bigl[ 
 \Bigl( \rho_\rmii{E} \Bigr)_\rmii{QCD}
 -      \Bigl( \rho_\rmii{E} \Bigr)_\rmii{HTL} \Bigr]_\rmii{naive}
 + \Bigl( \rho_\rmii{E} \Bigr)_\rmii{IR, resum} 
 + \Bigl[ \Bigl( \rho_\rmii{E} \Bigr)_\rmii{HTL}
 -      \Bigl( \rho_\rmii{E} \Bigr)_\rmii{IR}
   \Bigl]_\rmii{resum} 
 \nn
 & = & 
 \underbrace{\Bigl( \rho_\rmii{E} \Bigr)_\rmii{QCD, naive}}
 + \underbrace{\Bigl( \rho_\rmii{E} \Bigr)_\rmii{IR, resum} 
 - \Bigl( \rho_\rmii{E} \Bigr)_\rmii{HTL, naive}}
 + \underbrace{\Bigl( \rho_\rmii{E} \Bigr)_\rmii{HTL, resum} - 
 \Bigl( \rho_\rmii{E} \Bigl)_\rmii{IR, resum}} 
 \;, 
 \nonumber  \hspace*{0.5cm} \\  
 & & 
 \hspace*{0.3cm}
 \mbox{\eq\nr{rho_hard}} \hspace*{2.5cm}
 \mbox{\eq\nr{rho_soft}} \hspace*{3.4cm}
 \mbox{\eq\nr{wdep}}
 \la{rho_full}
\ea
where we left out the arguments $\omega$ for brevity.
The term $\bigl( \rho_\rmii{E} \bigl)_\rmii{IR, resum}$, 
added and subtracted in \eq\nr{rho_full}, is {\em defined}
to be the HTL-resummed spectral function but computed only up to 
the leading (linear) order in a Taylor expansion around
$\omega = 0$. The rationale for this split-up is that 
all three structures are ultraviolet and infrared finite 
(i.e.\ contain no $1/\epsilon$ divergences) and that 
the last term, \eq\nr{wdep}, is readily available from 
the literature (cf.\ below).

The first term of \eq\nr{rho_full} 
is the sum of \eqs\nr{rhoE2}, \nr{rhoEbare}, \nr{ct}: 
\ba
 \Bigl[ \rho_\rmii{E}(\omega) \Bigr]_\rmii{QCD, naive} & = & 
 \frac{g^2 C_F \omega^3}{6\pi} \biggl\{ 1 + 
 \frac{g^2}{(4\pi)^2}\biggl[ - \Nf 
 \biggl( \fr23 \ln\frac{\bmu^2}{4\omega^2} + \frac{20}{9} \biggr)
 \nn & & \hspace*{3cm} +\; \Nc \biggl( 
 \frac{11}{3} \ln\frac{\bmu^2}{4\omega^2} 
 + \frac{149}{9} - \frac{8\pi^2}{3}
 \biggr) \biggr] \biggr\} 
 \nn & & \; 
 + 
 \frac{g^2 C_F}{6\pi} \frac{g^2}{2\pi^2} \biggl\{ 
 \nn 
 & & \; + \; 
 \Nf \int_0^\infty \! {\rm d} q \, \nF{}(q) 
 \biggl[
   \Bigl(q^2 +\frac{\omega^2}{2}\Bigr) \ln \left| \frac{q+w}{q-w} \right| 
   + q \omega \biggl( \ln\frac{|q^2 - \omega^2|}{\omega^2} - 1 \biggr)\biggr] 
 \nn  & & \; + \; 
 \Nc \int_0^\infty \! {\rm d} q \, \nB{}(q) 
 \biggl[
   (q^2 +2 \omega^2) \ln \left| \frac{q+w}{q-w} \right| 
   + q \omega \biggl( \ln\frac{|q^2 - \omega^2|}{\omega^2} - 1 \biggr)
  \nn & & \hspace*{3cm} +\;
      \frac{\omega^4}{q} 
     \mathbbm{P} \biggl( \frac{1}{q+\omega} \ln \frac{q + \omega}{\omega} 
     +   \frac{1}{q-\omega} \ln \frac{\omega}{|q - \omega|} \biggr) \biggr]
 \biggr\} \;. \hspace*{0.7cm}
 \nn \la{rho_hard}
\ea
We stress that this spectral function contains no ``transport peak'' 
$\sim \omega\delta(\omega)$ (in fact none of the master sum-integrals
appearing in our computation have such terms). In addition, the function
$\tilde G_\rmii{E}(\omega_n)$, the cut of which
determines $\rho_\rmii{E}(\omega)$, 
contains no term constant in $\omega_n$, which would correspond 
to a contact term in the Euclidean correlator $G_\rmii{E}(\tau)$. 
Therefore $\rho_\rmii{E}(\omega)$ contains all the information 
characterizing the correlator in an ``unproblematic'' form
(see ref.~\cite{hbm_n} for a recent discussion of this issue
in a context where complications do appear).

The second term of \eq\nr{rho_full} is a combination of 
\eqs\nr{rho2_HTL_unres}, \nr{rhoE_HTL_unres}, \nr{rhoE_IR_resum}: 
\be
 \Bigl[ \rho_\rmii{E}(\omega) \Bigr]_\rmii{IR, resum} 
 - \Bigl[ \rho_\rmii{E}(\omega) \Bigr]_\rmii{HTL, naive} 
 =  \frac{g^2 C_F}{6\pi}\biggl[ - \omega^3 + 
     \fr12 \omega\mE^2 
     \biggl(\ln\frac{2\omega}{\mE} - 1 \biggr)
     \biggr]
 \;, 
 \la{rho_soft}
\ee
where $\mE$ is the Debye mass parameter, 
\be
  \mE^2 = \biggl( \frac{\Nf}{6} + \frac{\CA}{3} \biggr)\, g^2 T^2  
  \;. \la{mE}
\ee 
The term of $\rmO(g^2)$ cancels in the sum of \eqs\nr{rho_hard}, 
\nr{rho_soft}; this term 
is contained in a modified form within \eq\nr{wdep}
(cf.\ \eq\nr{HTL_asy}).
It is also interesting to note that, 
inserting $\mE^2$ as expressed in \eq\nr{mmE}, the next-to-leading
order contribution in \eq\nr{rho_soft} can be rewritten as 
\be
 \Bigl[ \rho_\rmii{E}^{(4)}(\omega) \Bigr]_\rmii{IR, resum} 
 - \Bigl[ \rho_\rmii{E}^{(4)}(\omega) \Bigr]_\rmii{HTL, naive} 
 =  \frac{g^2 C_F}{6\pi} \frac{g^2}{2\pi^2}
    \int_0^\infty \!\! {\rm d}q \, 
    \Bigl[ \Nf \, \nF{}(q) + \Nc \, \nB{}(q) \Bigr] 
    q \omega 
     \biggl(\ln\frac{4\omega^2}{\mE^2} - 2 \biggr)
 \;, 
 \la{rho_soft_2}
\ee
which nicely combines with some of the terms in \eq\nr{rho_hard}.

The remaining ingredient, 
the HTL-resummed contribution beyond the linear term in $\omega$, i.e.\ 
$
  ( \rho_\rmii{E} )_\rmii{HTL, resum}  - 
  ( \rho_\rmii{E} )_\rmii{IR, resum}
$, 
was determined in ref.~\cite{eucl}. 
With a minor change of notation, \eq(3.8) of ref.~\cite{eucl} 
can be expressed as
\ba
 & & \hspace*{-1cm}
  \Bigl[ \rho_\rmii{E}^{(4)}(\omega) \Bigr]_\rmii{HTL, resum}  - 
  \Bigl[ \rho_\rmii{E}^{(4)}(\omega) \Bigr]_\rmii{IR, resum}
  =  
 \frac{g^2 C_F \mE^2 \,\omega}{6\pi}  \times
 \biggl\{ \nn 
 & & 
 \int_{\hat\omega}^{\infty} \!  \frac{{\rm d}\hat k \, \hat k}{2} \, 
 \frac{\hat\omega^2
 \Bigl( 1 - \frac{\hat\omega^2}{\hat k^2}\Bigr)}
 {
  \Bigl( 
    \hat k^2 - \hat\omega^2 + \fr12 
     \Bigl[ 
       \frac{\hat\omega^2}{\hat k^2} + 
       \frac{\hat\omega}{2\hat k} 
       \Bigl( 1 - \frac{\hat\omega^2}{\hat k^2}\Bigr) 
       \ln\frac{\hat k + \hat\omega}{\hat k - \hat\omega}
     \Bigr]
  \Bigr)^2
 + \Bigl( 
     \frac{\hat\omega\pi}{4\hat k}
  \Bigr)^2
       \Bigl( 1 - \frac{\hat\omega^2}{\hat k^2}\Bigr)^2 
 } \nn 
 & & + \; 
 \int_{0}^{\infty} \! \frac{ {\rm d}\hat k \, \hat k^3}{2} 
 \biggl[ 
 \frac{
 \theta(\hat k - \hat\omega) 
 }
 {
  \Bigl( 
    \hat k^2 
     + 1 - 
       \frac{\hat\omega}{2\hat k} 
       \ln\frac{\hat k + \hat\omega}{\hat k - \hat\omega}
       \Bigr)^2
 + \Bigl( 
     \frac{\hat\omega\pi}{2\hat k}
  \Bigr)^2
 } 
 - \frac{1}{(\hat k^2 + 1)^2}
 \biggr]
 \nn 
 & & + \; 
 \left. 
   \frac{ {2 \hat\omega} \hat k_T^3 (\hat\omega^2 - \hat k_T^2)}
   {|3(\hat k_T^2 - \hat\omega^2)^2 -\hat\omega^2|}
 \right|_{\hat k_T^2 - \hat\omega^2 + \fr12
      [\frac{\hat\omega^2}{\hat k_T^2}+
        \frac{\hat\omega}{2\hat k_T} 
       ( 1 - \frac{\hat\omega^2}{\hat k_T^2} ) 
       \ln\frac{\hat\omega + \hat k_T }{\hat\omega - \hat k_T}
       ] \; = \; 0 }
\nn 
 & & + \; 
 \left. 
   \frac{\hat k_E^3 (\hat\omega^2 - \hat k_E^2)}
   { \hat\omega |3(\hat k_E^2 - \hat\omega^2) + 1|}
 \right|_{\hat k_E^2 + 1 -
        \frac{\hat\omega}{2\hat k_E} 
       \ln\frac{\hat\omega + \hat k_E }{\hat\omega - \hat k_E} \; = \; 0 }
 \quad \biggr\} \;, \la{wdep}
\ea
where $\hat\omega \equiv \omega/\mE$ and $\hat k \equiv k/\mE$, 
and the expression within the curly brackets is dimensionless. 

For $\hat\omega\gg 1$ it can be checked,
perhaps most easily numerically, that  
the expression within the curly brackets behaves as
\be
 \{ \ldots \}\; \stackrel{\hat\omega\gg 1}{=} \; 
 \hat\omega^2 + \fr12 \biggl( \ln\frac{1}{2 \hat\omega} + 1 \biggr)
 + \rmO\biggl( \frac{1}{\hat\omega^2} \biggr) 
 \;.  \la{HTL_asy}
\ee 
This cancels against
\eq\nr{rho_soft}, so that for $\omega\gg \mE$
only the naive QCD contribution of \eq\nr{rho_hard} is left over, 
as must be the case. 

%
\subsection{Ultraviolet asymptotics}
\la{ss:uv}

An important application of the general result of \se\ref{ss:full} is 
that it determines the asymptotic 
($\omega\gg \{ T, \mE, \Lambdamsbar \}$)
behaviour of the colour-electric spectral function. In fact, 
as explained in connection with \eq\nr{HTL_asy}, for large frequencies
$\omega\gg \mE$ 
we can make directly use of the unresummed expression, \eq\nr{rho_hard}.
Expanding the integrand of \eq\nr{rho_hard} 
in $q/\omega$ and making use of 
\be
 \int_0^\infty \! \frac{{\rm d}x \, x^3}{e^x -1} = \frac{\pi^4}{15} 
 \;, \quad
 \int_0^\infty \! \frac{{\rm d}x \, x^3}{e^x +1} = \frac{7\pi^4}{120} 
 \;, 
\ee
it can be seen that the thermal part of the spectral function 
disappears at large frequencies: 
\be
 \rho_\rmii{QCD}(\omega) \stackrel{\omega\gg T}{=} 
 \Bigl[ \rho_\rmii{QCD}(\omega) \Bigr]_{T=0}
 + \frac{g^4C_F}{6\pi} \frac{\pi^2T^4}{180\omega}
 ({7\Nf - 11\Nc})
 \;. 
 \la{asympto}
\ee
Such a power-law decay of the thermal correction 
at $\omega\gg T$ is in accordance with the general results of
ref.~\cite{sch2}, although it is not clear to us whether
a precise relation to the Operator Product Expansion
as used in ref.~\cite{sch2} can be established
for our non-local correlator.
The term denoted by ``$T=0$'' in \eq\nr{asympto}, on the other hand, 
is temperature independent, 
and given by the first two rows of \eq\nr{rho_hard}:
\ba
 \Bigl[ \rho_\rmii{QCD}(\omega) \Bigr]_{T=0} & = & 
 \frac{g^2 C_F \omega^3}{6\pi} \biggl\{ 1 + 
 \frac{g^2}{(4\pi)^2}\biggl[ - \Nf 
 \biggl( \fr23 \ln\frac{\bmu^2}{4\omega^2} + \frac{20}{9} \biggr)
 \nn & & \hspace*{3cm} +\; \Nc \biggl( 
 \frac{11}{3} \ln\frac{\bmu^2}{4\omega^2} 
 + \frac{149}{9} - \frac{8\pi^2}{3}
 \biggr) \biggr] \biggr\} + \rmO(g^6) 
 \;. \hspace*{0.7cm} \la{rhoASY}
\ea
A rough estimate can be obtained by setting $\bmu\approx 2\omega$ 
and inserting the 1-loop running coupling, 
\be
 \Bigl[ \frac{\rho_\rmii{QCD}(\omega)}{\omega^3} \Bigr]_{T=0}  
  \simeq  
 \frac{4\pi C_F}{(11\Nc - 2\Nf) \ln(2\omega/\Lambdamsbar)}
 \biggl\{ 1 - 
 \frac{(24\pi^2-149) \Nc + 20 \Nf }
 {6 (11\Nc - 2\Nf) \ln(2\omega/\Lambdamsbar) }
 \biggr\} 
 \;. \la{rough}
\ee
Numerical evaluations are discussed in \se\ref{ss:num}.

It is perhaps appropriate to stress that the first omitted
term in \eq\nr{HTL_asy} (which is not explicitly cancelled within
our computation) leads to a contribution of the type
$\delta\rho_\rmii{E}\sim g^2 \mE^4/\omega$, which is parametrically
of higher order than the $T$-dependent part of \eq\nr{asympto}. 
Hence, at $\rmO(\alpha_s^2)$,  ultraviolet asymptotics is indeed
completely determined by hard modes. This is again in accordance 
with the general analysis of ref.~\cite{sch2}

%
\subsection{Infrared asymptotics}
\la{ss:ir}

We next discuss the infrared ($\omega \ll \mE$) behaviour of the 
spectral function, in the range $T \gg \Lambdamsbar$. Although we have 
nothing to add to the existing results in this 
regime~\cite{mt,chm}, it is 
comforting to crosscheck that we can at least reproduce the expression
at $\rmO(\alpha_s^2)$. 
Since the computation of ref.~\cite{mt} was carried out in Coulomb 
gauge whereas we have described a covariant gauge computation, 
this also serves as a further
cross-check of gauge independence.

In the regime $\omega \ll \mE$ the ``$T=0$'' contribution 
(\eq\nr{rhoASY})
is subdominant compared with the thermal one, 
which is of $\rmO(g^2 \mE^2 \omega)$.
Expanding the thermal integrand of \eq\nr{rho_hard}
in $\omega/q$ and making use of 
\ba
 && \int_0^\infty \! \frac{{\rm d}x \, x}{e^x -1} = \frac{\pi^2}{6}
 \;, \quad
 \int_0^\infty \! \frac{{\rm d}x \, x \ln x }{e^x -1} = 
 \frac{\pi^2}{6} \biggl[ 1 - \gammaE + \frac{\zeta'(2)}{\zeta(2)} \biggr] 
 \;, \la{nB_asy_int} \\
 && \int_0^\infty \! \frac{{\rm d}x \, x}{e^x +1} = \frac{\pi^2}{12}
 \;, \quad
 \int_0^\infty \! \frac{{\rm d}x \, x \ln x }{e^x +1} = 
 \frac{\pi^2}{12} 
 \biggl[ \ln2+ 1 - \gammaE + \frac{\zeta'(2)}{\zeta(2)} \biggr] 
 \;,  
\ea
we obtain the infrared behaviour of the unresummed QCD result, expressed
now in terms of the momentum diffusion coefficient of \eq\nr{kappa_def_3}: 
\ba
 & & \hspace*{-1cm}
 {[\kappa]}_\rmi{QCD, naive}  =  
 \lim_{\omega\to 0} 
 \biggl[ \frac{2 T \rho_\rmii{E}(\omega)}{\omega} \biggr]_\rmi{QCD, naive}
 \nn 
 & \approx & 
 \frac{g^4 C_F T^3}{6\pi} \biggl\{ \frac{\Nc }{3}
 \biggl[
   \ln\frac{T}{\omega} + \fr32 - \gammaE +  \frac{\zeta'(2)}{\zeta(2)}
 \biggr] 
 + \frac{\Nf }{6}
 \biggl[
   \ln\frac{2T}{\omega} + \fr32 - \gammaE +  \frac{\zeta'(2)}{\zeta(2)}
 \biggr] \biggr\} 
 \;. \la{kappa_hard}
\ea
The symbol ``$\approx$'' signals that the logarithmically 
divergent term has been kept finite on the right-hand side. 
Similarly, from \eq\nr{rho_soft}, the HTL contribution to $\kappa$ is 
\ba
 {[\kappa]}_\rmi{HTL, resum} 
  - 
 {[\kappa]}_\rmi{HTL, naive} 
 & \approx & 
 \frac{g^2 C_F T}{6\pi} \mE^2 
 \biggl[ 
   \ln \frac{2\omega}{\mE} - 1
 \biggr]
 \;. \la{kappa_soft}
\ea
Summing together \eqs\nr{kappa_hard} and \nr{kappa_soft}, 
we reproduce the result of ref.~\cite{mt}: 
\be
 \kappa = \frac{g^2 C_F T}{6\pi} \mE^2 
 \biggl(\ln\frac{2 T}{\mE} + 
 \fr12 - \gamma_\rmii{E} + \frac{\zeta'(2)}{\zeta(2)}
 + \frac{\Nf \ln 2}{2\Nc + \Nf} \biggr)
 \biggl( 1 + \rmO(g) \biggr)
 \;. \la{kappa_lo}
\ee
As indicated, corrections to this expression start already
at $\rmO(g)$, and have been determined in ref.~\cite{chm}, 
but in the infrared regime
these go beyond the accuracy of our computation.

We note that if $g$ is not small ($T/\mE$ is not large), 
the expression in \eq\nr{kappa_lo} can become negative. This 
unphysical behaviour is a result of extrapolating the asymptotic
weak-coupling expression beyond its range of applicability; for
$\Nc = 3, \Nf = 0$, the problem occurs already at $g\gsim 1.05$ and 
is another reflection of the exceptionally poor convergence of the 
weak-coupling expansion of $\kappa$ (see also ref.~\cite{chm}).
Given that the unresummed expression (\eq\nr{kappa_hard})
does stay positive and in fact diverges as $\omega\to 0$, 
the HTL-resummation (\eq\nr{kappa_soft}) in some sense ``over-corrects'' 
the result in this regime.

%
\subsection{Numerical evaluation}
\la{ss:num}

In order to finally evaluate our result 
(\eq\nr{rho_hard} + \nr{rho_soft} + \nr{wdep}, cf.\ \eq\nr{rho_full})
numerically, we need 
to insert some value for the running coupling $g^2$.
It is only in the regime $\omega \gg T$ that two subsequent
orders with the same functional form are at our disposal, 
so that some kind of a scale optimization is possible; 
we can then define $\bmu_\rmi{opt}$ for $g^2$ from 
the ``fastest apparent convergence'' or ``principal of minimal
sensitivity'' criterion based on \eq\nr{rhoASY}: 
\be
 \ln(\bmu_\rmi{opt($\omega$)}) \equiv \ln(2 \omega) 
 + \frac{(24\pi^2-149)\Nc+20\Nf}{6(11\Nc-2\Nf)}
 \;, \la{muopt_w}
\ee
a structure indirectly also visible in \eq\nr{rough}. 
In the infrared regime $\omega\ll T$ such an exercise 
is not possible but we nevertheless need a value for $g^2$; 
we choose it from a context 
in which a next-to-leading order computation does exist, 
namely that of ``EQCD'' (cf.\ ref.~\cite{adjoint} 
and references therein):\footnote{%
  For EQCD even a NNLO computation 
  exists~\cite{gE2} but that goes beyond the accuracy of the 
  present study.  
  } 
\be
 \ln(\bmu_\rmi{opt($T$)}) \equiv \ln(4\pi T) - \gammaE - 
 \frac{\Nc-8\ln2\, \Nf}{2(11\Nc-2\Nf)}
 \;. \la{muopt_T}
\ee 
Comparing \eqs\nr{muopt_w} and \nr{muopt_T} we see that we can 
switch from one regime to another at $\omega\sim T$, 
for instance at $\omega \approx 0.8903T$ for $\Nf = 0$.
In order to get an error band for the uncertainty related
to the choice of $g^2$, we use the 
2-loop $g^2$,  with the scale parameter 
varied in the range  $(0.5 \ldots 2.0)\times \bmu_\rmi{opt}$.
Since the Debye mass parameter plays no role in the range
$\omega\gg T$ that we are mostly interested in, we fix it 
in the simplest way imaginable, through \eq\nr{mE} with the gauge coupling
evaluated as described above.

\begin{figure}[t]


\centerline{%
\epsfysize=7.5cm\epsfbox{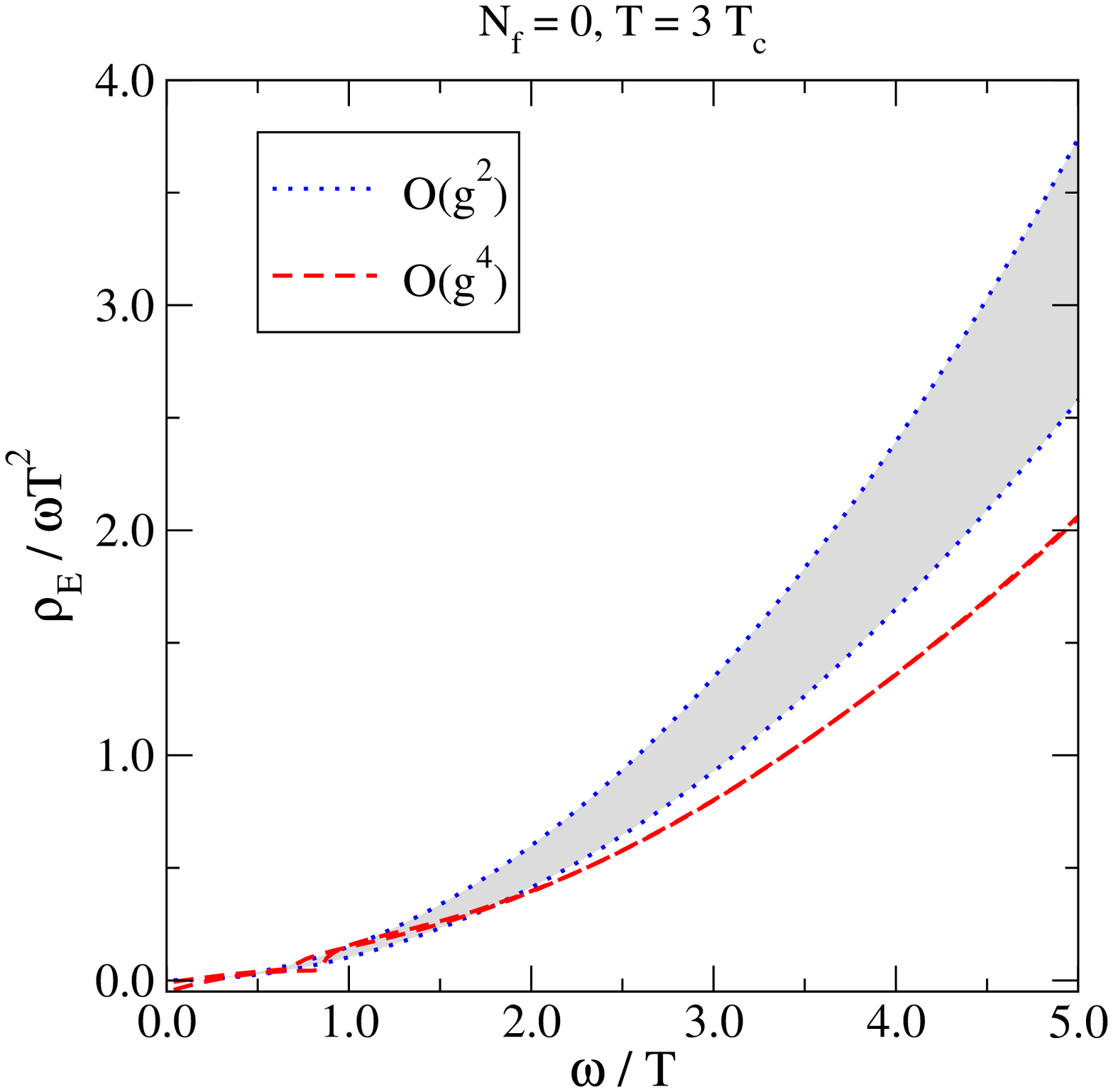}%
~~\epsfysize=7.5cm\epsfbox{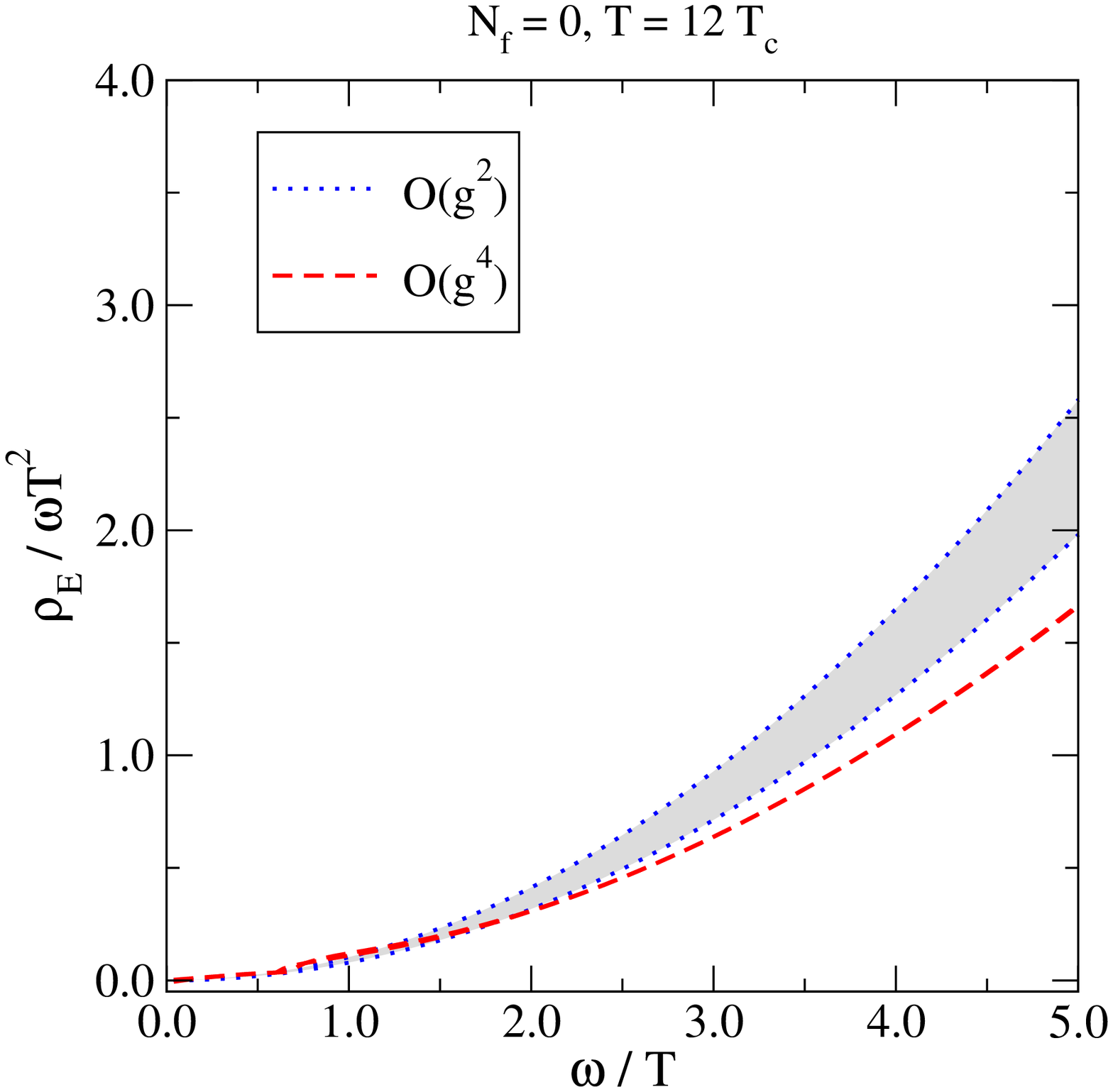}%
}

\caption[a]{\small 
  A numerical evaluation of $\rho_\rmii{E}(\omega)$, \eq\nr{rho_full},
  in units of  $\omega T^2$, for $\Nc = 3, \Nf = 0$, 
  $T = 3.75 \Lambdamsbar$ (left)
  and 
  $T = 15 \Lambdamsbar$ (right), 
  corresponding to   
  $T \approx 3 \Tc$ and
  $T \approx 12 \Tc$, respectively.
  The gauge coupling has been fixed as explained 
  around \eqs\nr{muopt_w}, \nr{muopt_T}, 
  and the error bar reflects the corresponding uncertainty.
  (In the $\rmO(g^2)$ result the ``optimal'' scale 
   is fixed to the thermal value of \eq\nr{muopt_T}, 
   i.e.\ does not change with the frequency.)
   A slightly negative intercept at $\omega\to 0$ 
   is an artifact of the truncated 
   weak-coupling expansion, cf.\ discussion below \eq\nr{kappa_lo}. 
 }
\la{fig:wdep}
\end{figure}

The outcome of a numerical evaluation of \eq\nr{rho_full} 
is plotted in \fig\ref{fig:wdep}, in units of $\omega T^2$.
We observe that because of the running of the gauge coupling
the next-to-leading order result falls below the leading-order one
in the ultraviolet domain, $\omega \gg T$. 
The dependence on the scale choice is also drastically reduced,
becoming (perhaps rather surprisingly) practically invisible 
as soon as $\omega \gsim T$.  

In the infrared domain, $\omega\ll T$, 
the next-to-leading order correction eventually overtakes
the leading-order result, and even the HTL-resummed expression
shows very slow convergence (cf.\ ref.~\cite{chm}). Nevertheless, 
\fig\ref{fig:wdep} does 
illustrate an important point: even at weak coupling there is 
no transport peak around the origin; 
rather $\rho_\rmii{E}(\omega)/\omega$ displays 
a relatively flat behavior at $\omega\lsim T$.
More precise studies of the infrared domain have so far been carried out
within classical lattice gauge theory~\cite{mink}
as well as within ${\mathcal N} = 4$
Super-Yang-Mills theory at 
infinite 't Hooft coupling~\cite{ct,ssg}.

\begin{figure}[t]


\centerline{%
\epsfysize=7.5cm\epsfbox{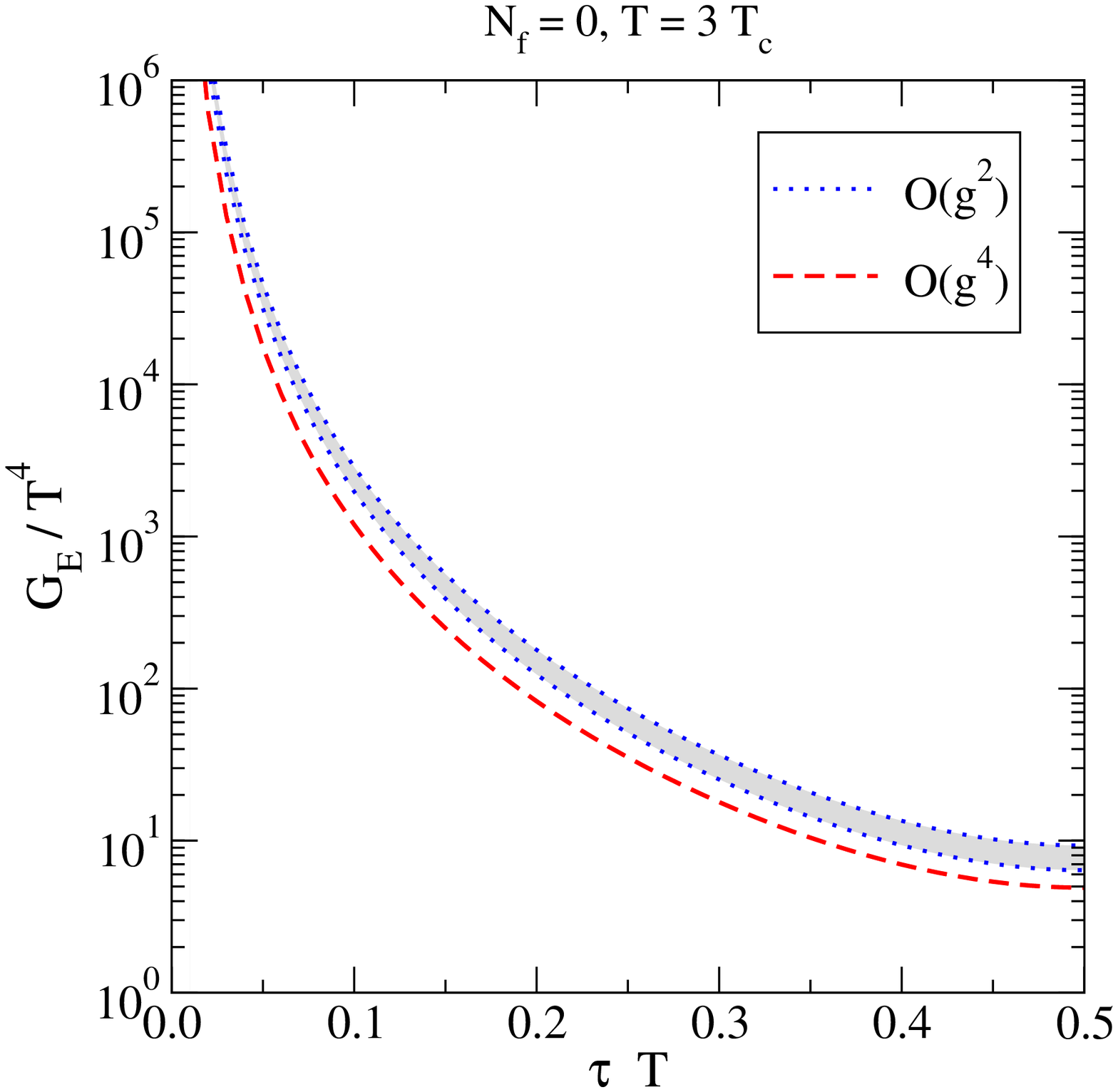}%
~~\epsfysize=7.5cm\epsfbox{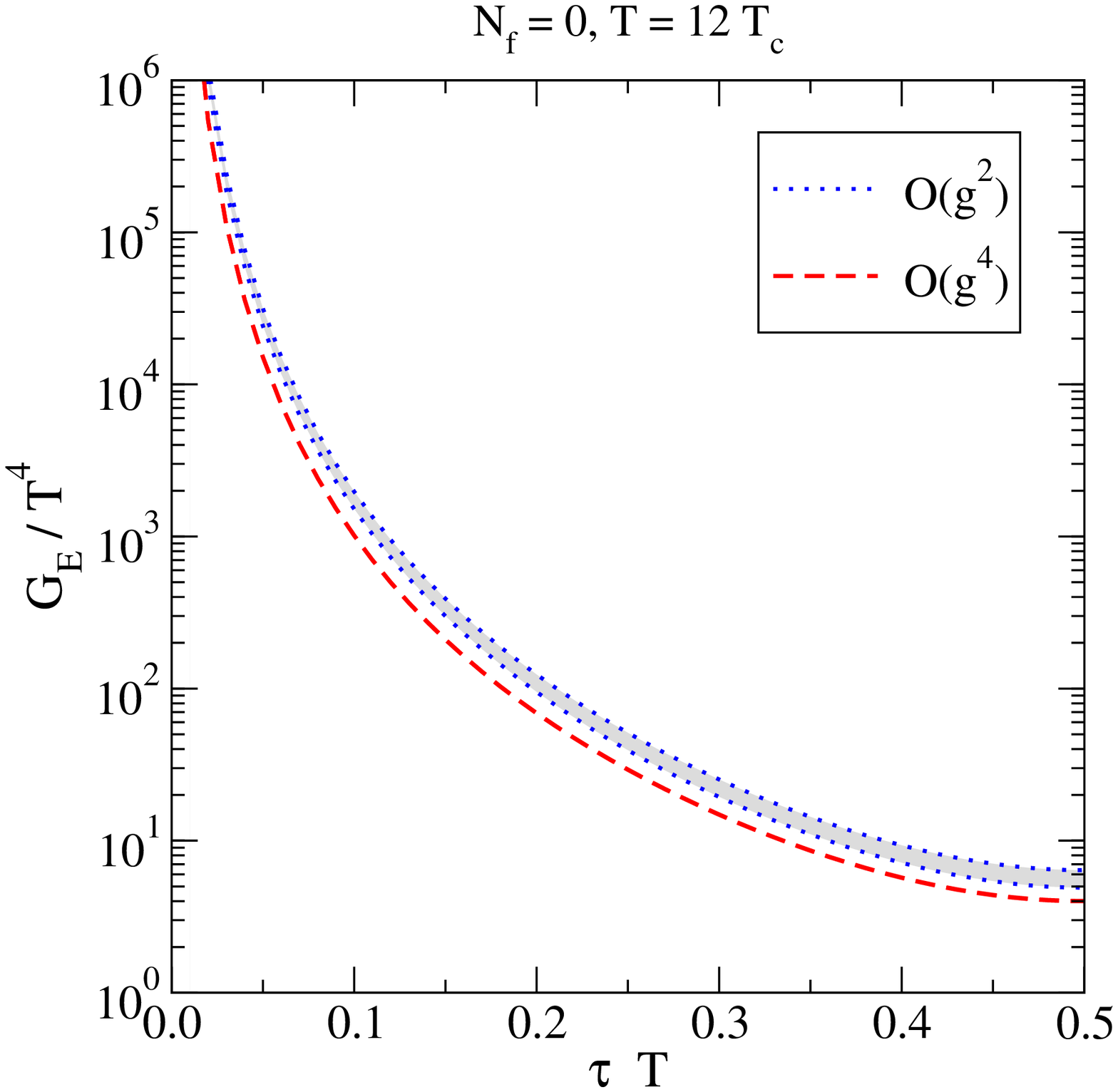}%
}

\caption[a]{\small 
  A numerical evaluation of $G_\rmii{E}(\tau)$,
  in units of  $T^4$, for $\Nc = 3, \Nf = 0$, 
  $T = 3.75 \Lambdamsbar$ (left)
  and 
  $T = 15 \Lambdamsbar$ (right), 
  corresponding to   
  $T \approx 3 \Tc$ and
  $T \approx 12 \Tc$, respectively.
  The gauge coupling has been fixed as explained 
  around \eqs\nr{muopt_w}, \nr{muopt_T}, 
  and the error bar reflects the corresponding uncertainty.
  (In the $\rmO(g^2)$ result the ``optimal'' scale 
   is fixed to the thermal value of \eq\nr{muopt_T}, 
   i.e.\ does not change with the frequency.)
  On the logarithmic scale used a clear 
  temperature-dependence is visible only around the middle
  of the interval, $\tau\, T \approx 0.5$.
 }
\la{fig:taudep}
\end{figure}

In \fig\ref{fig:taudep} we 
plot the Euclidean correlators $G_\rmii{E}(\tau)$
corresponding to the spectral function 
$\rho_\rmii{E}(\omega)$ of \fig\ref{fig:wdep}, obtained through
\eq\nr{int_rel}. The leading-order result can be given in 
a closed form~\cite{eucl},
\be
 G_\rmii{E}^{(2)}(\tau)  =  g^2 C_F\, \pi^2 T^4 \left[
 \frac{\cos^2(\pi \tau T)}{\sin^4(\pi \tau T)}
 +\frac{1}{3\sin^2(\pi \tau T)} \right]
 \;, \la{Gtau_free}
\ee
whereas the next-to-leading order correction 
$G_\rmii{E}^{(4)}$ has been evaluated numerically.
The $\omega$-dependent difference of $\rho_\rmii{E}$ at 
$\rmO(g^2)$ and $\rmO(g^4)$ in \fig\ref{fig:wdep}
has converted to a practically 
$\tau$-independent shift on the logarithmic scale of \fig\ref{fig:taudep}.

We note that, parametrically, the range 
$\omega \lsim \mE \ll 2T$ gives 
a contribution to \eq\nr{int_rel} of magnitude
\be
 \Bigl[ \delta G_\rmii{E}^{(4)}(\tau) \Bigr]_\rmii{HTL} 
 \sim 
 \int_0^{\mE} \! \frac{{\rm d}\omega}{\pi} 
 \frac{T \rho_\rmii{E}(\omega)}{\omega}
 \sim 
 g^2 \mE^3 T \sim \rmO(g^5 T^4)
 \;, 
\ee
cf.\ \eq\nr{kappa_soft}, which is 
formally of higher order than the ultraviolet contribution, 
$\rmO(g^4T^4)$. 
This is also reflected through 
the fact the infrared divergence of \eq\nr{kappa_hard} is integrable 
at small $\omega$. Nevertheless, the next ultraviolet contributions 
are of higher order still, 
$\rmO(g^6 T^4)$, so we can already meaningfully include even 
the infrared part of $\rho_\rmii{E}(\omega)$ in our numerical 
evaluation of  $G_\rmii{E}$.
In practice, of course, the scale $\mE$ is not  
smaller than $\sim 2T$ at phenomenologically interesting
temperatures, and the infrared contribution should not
be substantially suppressed.

%
\section{Colour-electric correlator in the strong-coupling expansion}
\la{se:strong}

In the confined (low-temperature) phase of pure SU($\Nc$) gauge 
theory, the Euclidean correlator of \eq\nr{GE_def} looks singular, 
because the expectation value in the denominator vanishes. We wish
to demonstrate that despite this appearance the ratio could 
actually remain finite, if defined through a suitable limiting
procedure by explicitly breaking the Z($\Nc$) center symmetry 
(an obvious possibility would be the addition of dynamical quarks). 
As a tool we use the lattice strong-coupling expansion, 
which converges (deep enough) in the confined phase. The result
also yields an interesting functional dependence on $\tau$, which 
can be contrasted with the weak-coupling result of \fig\ref{fig:taudep}.

Defining the strong-coupling expansion requires the use of 
lattice regularization, and hence also a discretization of
the correlator of \eq\nr{GE_def}. The discretization
is obviously not unique; we choose the possibility 
proposed in \eq(4.2) of ref.~\cite{eucl}, {\em viz.} 
\ba
 && \hspace*{-0.5cm} \LattGE \nn
 \la{GE_lattice}
\ea
where $a$ is the lattice spacing and 
we have used a graphical notation with lines indicating
parallel transporters (for more details see ref.~\cite{eucl}).
Let us also define the auxiliary correlator
\ba
 && \hspace*{-0.5cm} \LattC 
 \la{C_lattice}
\ea
in which the vertical lines have the length of one lattice spacing.
If $C(0)\neq 0$, we can then write
\be
 G_\rmii{E}(\tau) = \frac{C(\tau-a) + C(\tau+a) - 2 C(\tau)}{a^4 C(0)}
 \;. \la{GE_C_rel}
\ee
Note that this expression makes sense only for $\tau \ge a$.

In order to guarantee that $C(0)$ is non-zero, we add a small 
``source term'' to the (Wilson) action, thus breaking the Z($\Nc$)
symmetry explicitly. We believe that the precise from of this regulator
does not matter, and in practice we choose the simplest possibility, 
adding to the action just the volume sum of the Polyakov loop itself, 
multiplied by a small ``magnetic field'', to be denoted by $\kappa$.
Then (if $N_\tau \ge 4$), 
\be
 C(\tau) = \kappa \Bigl[u^{\frac{\tau}{a}}
 + u^{\frac{\beta-\tau}{a}} \Bigr] 
 \Bigl( 1 + \rmO(u^{4}) \Bigr) + \rmO(\kappa^2)
 \;,  \la{C_res}
\ee
where $N_\tau \equiv \beta/a$ and $u$ is a variable naturally
arising in the character expansion 
(for details see, e.g., ref.~\cite{mm}), 
\be
 u \equiv \frac{1}{\Nc} \tr  \langle\!\langle P_{ij} \rangle\!\rangle
 \;. \la{u_def}
\ee
Here $P_{ij}$ denotes an elementary plaquette in the $(i,j)$-plane, 
and the expectation value is defined as 
$
 \langle\!\langle P_{ij} \rangle\!\rangle \equiv
 \int {\rm d} P_{ij} \, P_{ij} 
 \exp(\frac{\beta_\rmii{G}}{2 \Nc} \tr [P_{ij}+P_{ij}^\dagger] ) /
 \int {\rm d} P_{ij} \,  
 \exp(\frac{\beta_\rmii{G}}{2 \Nc} \tr [P_{ij}+P_{ij}^\dagger] )
$. 
The strong-coupling expansion is an expansion in a small 
Wilson parameter $\beta_\rmii{G}$, which means that the variable $u$
should also be small: for $\Nc \ge 3$, 
\be
 u = \frac{\beta_\rmii{G}}{2 \Nc^2} + \rmO(\beta_\rmii{G}^2)
 \;.
\ee
Incidentally, 
we have also worked out the correction of $\rmO(u^4)$ to \eq\nr{C_res}; 
it amounts to the replacement $u^x\to (1 + 4 x u^4)u^x$, 
with $x=\frac{\tau}{a}$ and $x=\frac{\beta-\tau}{a}$,   
but this does not change the qualitative behaviour so 
we do not elaborate on further details here.

\begin{figure}[t]


\centerline{%
 \epsfysize=9.0cm\epsfbox{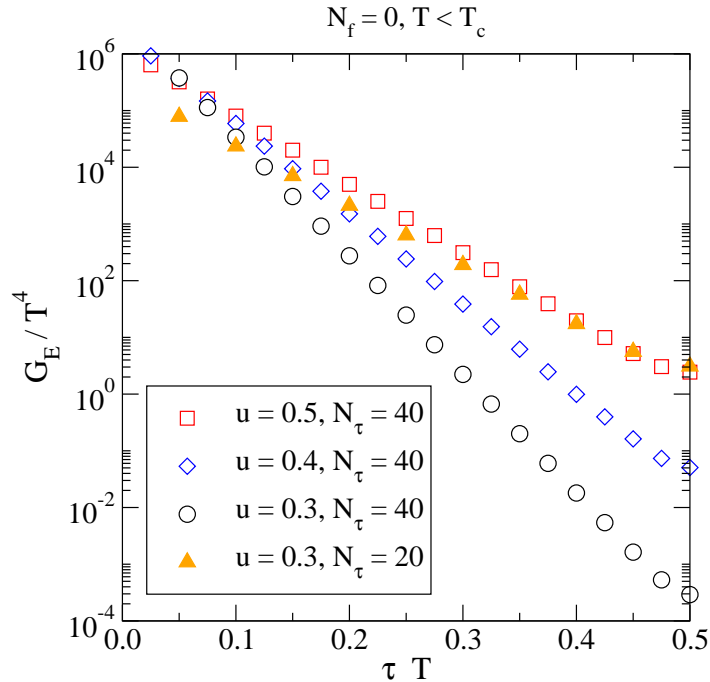}%
}

\caption[a]{\small 
  A numerical evaluation of the strong-coupling result 
  for $G_\rmii{E}(\tau)$, \eqs\nr{GE_C_rel}, \nr{C_res},  
  in units of  $T^4$. For a fixed $N_\tau$, 
  smaller values of $u$ correspond to lower temperatures.  
  The qualitative behaviour can be compared with that 
  in \fig\ref{fig:taudep}; the exponential
  suppression around the middle of the interval becomes 
  very strong at low temperatures (note that the scales 
  are different in the two figures). 
 }
\la{fig:strongcoupling}
\end{figure}

Inserting \eq\nr{C_res} into \eq\nr{GE_C_rel}, we note that the 
magnetic field $\kappa$ drops out, and a finite result is left over. 
Moreover, inserting some small value for $u$ ($0 < u \ll 1$), 
we obtain a function which is not unlike those in \fig\ref{fig:taudep};
examples are shown in \fig\ref{fig:strongcoupling}. The main
differences are that at very low temperatures (very small $u$), 
the exponential suppression around the middle of the interval
is much stronger than in the deconfined phase; and that there is 
little curvature in the plots (i.e.\ only a single exponential).
Taking a discrete Fourier transform and 
subsequently {\em naively} applying \eq\nr{disc} to it, 
leads to a spectral function with 
$\delta$-function peaks at $\omega = \pm \frac{1}{a} \ln u$
but no other structure. This behaviour is completely
different from those in \fig\ref{fig:wdep}; the difference 
demonstrates once again that a high resolution is 
needed in the determination of $G_\rmii{E}(\tau)$, in order
for us to be able to extract the correct non-trivial features 
of the corresponding spectral function.

In a practical lattice measurement, the numerator and 
the denominator of \eq\nr{GE_C_rel} are evaluated separately. 
If the breaking of the Z($\Nc$) symmetry is only explicit 
(not spontaneous) and small, the signal is likely to be very noisy. 
This is not really a problem from the phenomenological point of view, 
though, since heavy quark diffusion appears anyways to be a physically 
meaningful concept only in the deconfined phase where Z($\Nc)$ is 
indeed spontaneously broken (at any non-zero lattice spacing).  

%
\section{Conclusions and outlook}
\la{se:concl}

The ``transport coefficient'' related to the correlator
of two colour-electric fields along a Polyakov loop determines
the momentum diffusion coefficient, $\kappa$, of a heavy quark near rest
with respect to a thermal medium. Among all the QCD transport coefficients
this is probably the most accessible one from the 
theoretical point of view; indeed it is the first 
for which a next-to-leading order weak-coupling expression
was obtained~\cite{chm}. Alas, as the weak-coupling 
expression itself~\cite{chm} as well as numerical studies
within classical lattice gauge theory~\cite{mink} show, 
the weak-coupling expansion converges very slowly at any temperature
relevant for heavy ion collision experiments, 
so a non-perturbative determination would be highly desirable. 

On the point of the non-perturbative determination, 
there is the salient feature 
that no transport peak exists in the colour-electric spectral function from 
which $\kappa$ is extracted, 
$\rho_\rmii{E}(\omega)$, 
even at weak coupling where spectral functions are in general 
more singular than in a strongly interacting system. 
The reason is that there is no (approximately) conserved charge
related to the colour-electric field strength.
Employing this theoretical insight as a constraint might help
to reduce the systematic uncertainties that are inherent 
to the practical recipes used for the analytic continuation from 
the measured Euclidean correlator to the desired $\rho_\rmii{E}(\omega)$.

On the other hand, the non-perturbative study 
is faced with the challenge that the
absolute magnitude of the correlator is needed for the 
determination of $\kappa$, which is obtained from the 
$\omega\to 0$ intercept of $2 T \rho_\rmii{E}(\omega)/\omega$.
This means that the renormalization of the colour-electric field 
entering the correlator plays an important role. 

In principle, 
the colour-electric field can be renormalized independently
of the particular observable considered
(perhaps in analogy with procedure for colour-magnetic fields
in ref.~\cite{gms}). However, its discretization is not unique, 
so it may be useful to 
have a way to ``crosscheck'' the normalization of the result 
directly with the data at hand. For this purpose our asymptotic
large-$\omega$ behaviour might turn out to be useful, since the 
weak-coupling result is relatively accurate at $\omega \gg T$ 
(at least within $\sim 10$~\%, 
which would at the current stage be perfectly sufficient 
as far as $\kappa$ is concerned). Our formula is strictly  
applicable only in the continuum limit; nevertheless, a practical matching
could already be attempted at some intermediate frequency $\omega$ and 
non-zero lattice spacing $a$, provided that the inequalities 
$T \ll \omega \ll 1/a$ can be satisfied.   
Of course, all of this assumes that some $\rho_\rmii{E}(\omega)$
can be approximately reconstructed from a measured Euclidean
correlator $G_\rmii{E}(\tau)$~\cite{recent}.

Apart from the spectral function (\fig\ref{fig:wdep}), 
we have also estimated the corresponding Euclidean correlator
(\figs\ref{fig:taudep}, \ref{fig:strongcoupling}) 
which can be directly compared with lattice simulations. 
The result appears to be most sensitive to thermal effects
around the middle of the Euclidean time interval, where 
its absolute value is very small, posing 
a challenge for numerical simulations. Nevertheless, with 
refined techniques a reasonable signal should ultimately be obtainable. 

To summarize, it is our (perhaps optimistic) 
hope that by combining our analytic work with
forthcoming numerical studies a qualitative and even
approximate quantitative understanding concerning the shape of the spectral
function $\rho_\rmii{E}(\omega)$ and in particular the intercept 
$\kappa = \lim_{\omega\to 0} 2 T \rho_\rmii{E}(\omega)/\omega$ 
can eventually be developed. 

%
\section*{Acknowledgements}

We are grateful to the BMBF for financial support under the project
{\em Heavy Quarks as a Bridge between Heavy Ion Collisions and QCD}.
M.L.\ was supported in part by the Yukawa International Program 
for Quark-Hadron Sciences at Yukawa Institute for 
Theoretical Physics, Kyoto University, Japan, and 
L.M.\ was supported through the Sofja Kovalevskaja program 
of the Alexander von Humboldt foundation. 
M.L.\ thanks A.~Francis and O.~Kaczmarek for interesting discussions
concerning the prospects for lattice simulations.

\newpage

\appendix
\renewcommand{\thesection}{Appendix~\Alph{section}}
\renewcommand{\thesubsection}{\Alph{section}.\arabic{subsection}}
\renewcommand{\theequation}{\Alph{section}.\arabic{equation}}

%
\section{Details of the unresummed computation}

%
\subsection{Matsubara sums, Fourier transform, spectral function}
\la{ss:mat}

We illustrate here the steps needed for the determination of 
the spectral function with the help of one of the basic structures 
appearing in the analysis, 
the function $\mathcal{I}_2$ defined in \eq\nr{I2_def}. 
It turns out that in the course of the computation the expression
splits into several parts, some of which may contain poles for 
specific spatial momenta; although such poles cancel at the end, 
it is important to ``regulate'' the intermediate expressions 
consistently so that no finite pieces go amiss. A convenient
way to accomplish this is to redefine $\mathcal{I}_2$ as 
\ba
 \mathcal{I}_2(\tau) & \equiv & \lim_{\lambda\to 0} \Tint{K,Q} 
 \frac{k^2 e^{i k_n\tau}}{[K^2+\lambda^2] Q^2(K-Q)^2}
 \;. \la{I2_def_new}
\ea
We have considered another regularization as well, writing
the denominator as $1/\{ K^2 [Q^2+\lambda^2][(K-Q)^2+\lambda^2] \}$, 
and checked that this leads to the same result in the limit 
$\lambda\to 0$. 

As a first step, we carry out the Matsubara sums over $k_n,q_n$. 
Defining
\be
 E_k \equiv \sqrt{k^2 + \lambda^2} \;, \quad
 E_q \equiv |\vec{q}| \;, \quad 
 E_{kq} \equiv |\vec{k-q}|
 \;, 
\ee
the sums can be written as 
\ba
 & & \hspace*{-2cm} T^2 \sum_{k_n,q_n} \frac{e^{ik_n\tau}}
 {(k_n^2 + E_k^2)(q_n^2 + E_q^2)[(k_n-q_n)^2 + E_{kq}^2]}
 \nn & = & 
 T^3 \sum_{k_n,q_n,r_n}  \frac{\beta \delta_{r_n+k_n-q_n} e^{ik_n\tau}}
 {(k_n^2 + E_k^2)(q_n^2 + E_q^2)(r_n^2 + E_{kq}^2)}
 \nn & = & 
 \int_0^\beta \! {\rm d}\sigma \, 
 T^3 \sum_{k_n,q_n,r_n} 
 \frac{e^{ik_n(\tau+\sigma)}}{k_n^2 + E_k^2} 
 \frac{e^{-iq_n\sigma}}{q_n^2 + E_q^2}
 \frac{e^{i r_n\sigma}}{r_n^2 + E_{kq}^2}
 \;, 
\ea
where in the last step we  used 
a representation of the (periodic) Kronecker delta, 
$
 \beta \delta_{t_n} = \int_0^{\beta}\!{\rm d}\sigma \, e^{i t_n\sigma}
$.
The sums have now factorized and can be carried out like in \eq\nr{sum}; 
care must be taken just in order to transform the time arguments to 
the interval $(0, \beta)$, which can be achieved for
\be
 \tilde G(\tau,E) \equiv T \sum_{q_n} \frac{e^{i q_n\tau}}{q_n^2 + E^2}
\ee
by making use of the symmetries
\ba
  \tilde G(-\tau,E) =  \tilde G(\tau + m\beta ,E) =  \tilde G(\tau,E)
  \;, \quad m\in\mathbbm{Z} \;. 
\ea
Since $\tau, \sigma$ only appear inside exponential 
functions in the result (cf.\ \eq\nr{sum}), 
the integral over $\sigma$ as well as 
the subsequent Fourier transform (\eq\nr{tildeGE}) are trivial. 
The discontinuity as in \eq\nr{disc} can finally be taken by first 
setting $\exp(\pm i\omega_n\beta) =1$ in order to remove 
terms growing exponentially at large $|\omega|$, and by
subsequently making use of 
\be
 \im \biggl[ \frac{1}{i\omega_n + \sum_i E_i} 
 \biggr]_{\omega_n\to -i[\omega+i0^+]}
 = -\pi \delta(\omega + \sum_i E_i)
 \;. \la{im_inv}
\ee
Note that by making use of this relation we are advised to replace
all other fractions by principal values 
($
\frac{1}{x+i0^+} = \mathbbm{P}(\frac{1}{x}) - i \pi \delta(x)
$).

Let us stress that all terms 
(in this as well as in other master sum-integrals)
contain $\omega_n$ exclusively 
within the structures of  \eq\nr{im_inv}, 
always with at least one non-zero $E_i$;  therefore 
no ``transport peaks'' $\sim\omega \delta(\omega)$ arise
in the analytic continuation. 
In addition, there are no $\omega_n$-independent terms, 
which would correspond to contact terms in configuration space. 

Denoting the result of these 
steps by $\tilde{\mathcal{I}}_2(\omega)$, we obtain
\ba
 \tilde{\mathcal{I}}_2(\omega) & = & \lim_{\lambda\to 0}
 \int_{\vec{k},\vec{q}} \frac{\pi k^2}{4E_qE_{kq}}
 \biggl\{
   \mathbbm{P}\biggl( \frac{1}{E_k^2 - \omega^2} \biggr) 
   \biggl[
     \delta(\omega - E_q - E_{kq}) (1+\nB{1}+\nB{2}) 
  \nn & & \hspace*{5.3cm} + \, \delta(\omega+E_q-E_{kq}) (\nB{1}-\nB{2})
  \nn & & \hspace*{5.3cm} + \, \delta(\omega-E_q+E_{kq}) (\nB{2}-\nB{1})
  \nn & & \hspace*{5.3cm} - \, \delta(\omega+E_q+E_{kq}) (1 + \nB{1}+\nB{2})
   \biggr] 
 \nn & & \hspace*{0.7cm}  + \, 
 \frac{\delta(\omega-E_k) - \delta(\omega+E_k)}{2 E_k}
 \, \mathbbm{P}\biggl[ 
   \frac{1}{E_k + E_q + E_{kq}} +    \frac{1}{-E_k + E_q + E_{kq}}
  \nn & & \hspace*{3.3cm} - \, 2 E_{kq}\, \nB{1}
  \biggl( \frac{1}{(E_k-E_q)^2 - E_{kq}^2} 
   + \frac{1}{(E_k+E_q)^2 - E_{kq}^2} \biggr)
  \nn & & \hspace*{3.3cm} - \, 2 E_{q}\, \nB{2}
  \biggl( \frac{1}{(E_k-E_{kq})^2 - E_{q}^2} 
   + \frac{1}{(E_k+E_{kq})^2 - E_{q}^2} \biggr)
 \biggr]
 \biggr\} 
 \;, \nn \la{I2_1}
\ea
where
\be
 \nB{1} \equiv \nB{}(E_q) \;, \quad
 \nB{2} \equiv \nB{}(E_{kq}) \;. 
\ee
The expression is obviously antisymmetric in $\omega\to -\omega$; 
in the following we restrict, without loss of generality, to $\omega > 0$.

In \eq\nr{I2_1}, two different qualitative structures can be observed. 
We refer to the first group, with the constraints 
$\delta(\omega - E_q - E_{kq})$ etc, as ``phase space integrals''; 
and to the second group, with the constraint $\delta(\omega - E_k)$, 
as ``factorized integrals''. We now proceed to discuss these two 
groups separately. The analysis parallels that in appendix~A
of ref.~\cite{nlo}.

%
\subsection{Phase space integrals}
\la{ss:ps}

The phase space integrals are ultraviolet finite so that we can set
$D\to 4$ from the outset. We fix the $z$-axis in the direction of $\vec{k}$
and carry out the integral over the directions of $\vec{q}$ by changing
integration variables: 
\be
 \int \! \frac{{\rm d}^3 \vec{q}}{(2\pi)^3} 
 \to 
 \frac{1}{4\pi^2 k} \int_0^\infty \! {\rm d}q \, q 
 \int_{|k-q|}^{k+q} \! {\rm d}E_{kq} \, E_{kq}
 \;. \la{meas1}
\ee
For $\omega > 0$, the following results are easily verified: 
\ba
 \int_{|k-q|}^{k+q} \! {\rm d}E_{kq} \, 
 \delta(\omega-E_q-E_{kq}) \, \phi(E_{kq})
 & = & \theta(\omega-k) \, 
       \theta\Bigl(\frac{\omega+k}{2} -q \Bigr)
       \theta\Bigl(q - \frac{\omega-k}{2} \Bigr)
       \phi(\omega - q)
 \;, \nn \la{ps1} \\ 
 \int_{|k-q|}^{k+q} \! {\rm d}E_{kq} \,
 \delta(\omega+E_q-E_{kq})\, \phi(E_{kq})
 & = & \theta(k-\omega) \, 
       \theta\Bigl(q - \frac{k-\omega}{2} \Bigr)
       \phi(q + \omega)
 \;,  \la{ps2} \\ 
  \int_{|k-q|}^{k+q} \! {\rm d}E_{kq} \,
 \delta(\omega-E_q+E_{kq})\, \phi(E_{kq})
 & = & \theta(k-\omega) \, 
       \theta\Bigl(q - \frac{k+\omega}{2} \Bigr)
       \phi(q - \omega)
 \;, \la{ps3}
\ea
where we also simplified the notation as $E_q \to q$.

To go further
let us start by considering the 
temperature-independent ``vacuum part'', 
to be denoted with $\delta_\rmii{ps,0} \tilde{\mathcal{I}}_2$:
\ba
 \delta_\rmii{ps,0}\, \tilde{\mathcal{I}}_2(\omega) & \equiv & 
 \lim_{\lambda\to 0}
 \int  \! \frac{{\rm d}^3\vec{k}}{(2\pi)^3} 
  \mathbbm{P}\biggl( \frac{\pi k^2 }{E_k^2 - \omega^2} \biggr) 
 \int  \! \frac{{\rm d}^3\vec{q}}{(2\pi)^3}
 \frac{\delta(\omega - E_q - E_{kq})}{4E_qE_{kq}}
 \;. \la{I2_ps0_1}
\ea
Changing integration
variables as in \eq\nr{meas1} and making use of \eq\nr{ps1}, we get
\ba
 \delta_\rmii{ps,0}\, \tilde{\mathcal{I}}_2(\omega) & = & 
 \lim_{\lambda\to 0} \frac{4\pi}{(2\pi)^3}
 \int_0^{\omega} \! {\rm d}k \, 
 \mathbbm{P}\biggl( \frac{\pi k^3 }{k^2 + \lambda^2 - \omega^2} \biggr)
 \frac{1}{(4\pi)^2}
 \int_{\frac{\omega-k}{2}}^{\frac{\omega+k}{2}}
 \! {\rm d} q 
 \nn & = & 
 \lim_{\lambda\to 0}
 \frac{1}{32\pi^3} 
 \int_0^{\omega} \! {\rm d}k \, 
 \mathbbm{P}\biggl( \frac{k^4 }{k^2 + \lambda^2 - \omega^2} \biggr)
 \nn & = & 
 \lim_{\lambda\to 0}
 \frac{\omega^3}{(4\pi)^3}
 \biggl( \ln\frac{\lambda^2}{4\omega^2} + \fr83 \biggr)
 \;. \la{I2_ps0_2}
\ea
We see how the phase space integration contains a divergence, 
originating from an ``on-shell'' configuration ($k\approx\omega$); 
this will be 
cancelled against a corresponding divergence in the factorized
vacuum integral, \eq\nr{I2_fz0_2}, such that in the complete
result the limit $\lambda\to 0$ can be taken.

Considering the thermal part, the substitution 
$\vec{q}\to\vec{k}-\vec{q}$, i.e.\ $E_q\leftrightarrow E_{kq}$, 
allows us to remove some redundancy, and the integral becomes
\ba
 & & \hspace*{-1.5cm} \delta_\rmii{ps,T}\, 
 \tilde{\mathcal{I}}_2(\omega) \equiv  
 \lim_{\lambda\to 0}
 \frac{4\pi}{(2\pi)^3} \int_0^\infty \! {\rm d}k\,  
 \mathbbm{P}\biggl[ \frac{\pi k^3 }{4(k^2 + \lambda^2 - \omega^2)} \biggr]
 \frac{1}{4\pi^2} \int_0^\infty \! {\rm d}q \, 
 \nn & \times & 
 \int_{|k-q|}^{k+q} \! {\rm d}E_{kq} \,
 \Bigl\{
  2\delta(\omega-q-E_{kq}) \nB{}(q) 
 + 2 \delta(\omega+q-E_{kq}) [\nB{}(q) - \nB{}(E_{kq}) ]   
 \Bigr\}
 \nn & = & 
 \lim_{\lambda\to 0}
 \frac{1}{16\pi^3}
 \int_0^\infty \!\!\! {\rm d}k\,  
 \mathbbm{P}\biggl( \frac{k^3}{k^2 + \lambda^2 - \omega^2} \biggr)
 \biggl\{
   \theta(k-\omega)\!\! \int_{\frac{k-\omega}{2}}^\frac{k+\omega}{2}
   \!\! {\rm d}q + 
   \theta(\omega-k)\!\! \int_{\frac{\omega-k}{2}}^\frac{\omega+k}{2}
   \!\! {\rm d}q
 \biggr\} \nB{}(q)
 \;, \nn \la{I2_psT_1}
\ea
where we made use of \eqs\nr{meas1}--\nr{ps2},  
and subsequently shifted $q+\omega\to q$ in one of the integrals, 
in order to always have $q$ as the argument of the Bose distribution. 
At this point it is advantageous to change the order of integration; 
the integration area can be illustrated as
$$
 \hspace*{5cm}
 \parbox[c]{60pt}{
 \begin{picture}(0,0)(120,60)
        \thicklines
	\put(0,0){\vector(1,0){120}}
	\put(0,0){\vector(0,1){60}}
	\put(0,30){\line(2,1){60}}
	\put(0,30){\line(2,-1){60}}
	\put(60,0){\line(2,1){60}}
        \put(-10,60){\makebox(0,0)[c]{${q}$}}
        \put(-10,30){\makebox(0,0)[c]{$\frac{\omega}{2}$}}
        \put(60,-10){\makebox(0,0)[c]{${\omega}$}}
        \put(120,-10){\makebox(0,0)[c]{${k}$}}
        \thinlines
        \put(60,60){\line(1,0){60}}
        \put(50,55){\line(1,0){70}}
        \put(40,50){\line(1,0){80}}
        \put(30,45){\line(1,0){90}}
        \put(20,40){\line(1,0){100}}
        \put(10,35){\line(1,0){110}}
        \put(0,30){\line(1,0){120}}
        \put(10,25){\line(1,0){100}}
        \put(20,20){\line(1,0){80}}
        \put(30,15){\line(1,0){60}}
        \put(40,10){\line(1,0){40}}
        \put(50,5){\line(1,0){20}}
 \end{picture}} 
 \vspace*{2.5cm}
$$
Thereby we obtain
\ba
 \delta_\rmii{ps,T}\, \tilde{\mathcal{I}}_2(\omega) & = & 
 \lim_{\lambda\to 0}
 \frac{1}{16\pi^3}
 \int_0^\infty \!\!\! {\rm d}q\, \nB{}(q) 
 \int_{|\omega - 2q|}^{\omega+2q} \!\!\! {\rm d}k\,  
 \mathbbm{P}\biggl( \frac{k^3}{k^2 + \lambda^2 - \omega^2} \biggr)
 \nn & = & 
 \frac{1}{16\pi^3}
 \int_0^\infty \!\!\! {\rm d}q\, \nB{}(q) 
 \biggl[ 4 q \omega + \frac{\omega^2}{2} 
   \ln\left| \frac{q+\omega}{q-\omega} \right| 
 \biggr]
 \;. \la{I2_psT_2}
\ea

%
\subsection{Factorized spatial integrals}
\la{ss:fz}

The factorized integrals are more delicate than the phase space ones in 
the sense that the $\vec{q}$-integral can contain ultraviolet divergences. 
In general this means that the integration measure is to be written as
\be
 \int \! \frac{{\rm d}^d\vec{q}}{(2\pi)^d} 
 \to 
 \frac{4}{(4\pi)^{\frac{d+1}{2}} \Gamma(\frac{d-1}{2})}
 \int_0^{\infty} \! {\rm d}q \, q^{d-1} 
 \int_{-1}^{+1} \! {\rm d}z \, (1-z^2)^{\frac{d-3}{2}}
 \;, \la{meas2}
\ee 
where $d\equiv D-1$ and $z = \vec{k}\cdot\vec{q}/kq$. 
If the integrand is independent of $z$, the $z$-integral yields
\be
 \int_{-1}^{+1} \! {\rm d}z \, (1-z^2)^{\frac{d-3}{2}}
 = \frac{\Gamma(\frac{1}{2})\Gamma(\frac{d-1}{2})}{\Gamma(\frac{d}{2})} 
 \;. \la{meas25}
\ee
If the $\vec{q}$-integral is finite, as is always the case
with thermal corrections, it is convenient to change
integration variables like in \eq\nr{meas1}, 
\be
 \int_{-1}^{+1} \! {\rm d}z \to \frac{1}{kq} 
 \int_{|k-q|}^{k+q} \! {\rm d}E_{kq} \, E_{kq}
 \;. \la{meas3}
\ee
Sometimes one is also in the lucky position that the $\vec{q}$-integral 
can be directly identified with a known vacuum integral (see below). 

Let us start by considering the 
temperature-independent ``vacuum part'', 
to be denoted by $\delta_\rmii{fz,0} \tilde{\mathcal{I}}_2$:
\ba
 \delta_\rmii{fz,0} \tilde{\mathcal{I}}_2(\omega) & \equiv & 
 \lim_{\lambda\to 0}
 \int  \! \frac{{\rm d}^d\vec{k}}{(2\pi)^d} 
  \frac{\pi k^2\delta(\omega-E_k)}{2 E_k} 
 \nn & & \; \times \, 
 \int  \! \frac{{\rm d}^d\vec{q}}{(2\pi)^d}
 \frac{1}{4E_qE_{kq}}
 \mathbbm{P}\biggl[ 
   \frac{1}{\omega + E_q + E_{kq}} +    \frac{1}{-\omega + E_q + E_{kq}}
 \biggr]
 \;. \la{I2_fz0_1}
\ea
Here we again restricted to $\omega > 0$.
Within the $\vec{k}$-integral the limit $\lambda\to 0$ can be 
immediately taken, so that $E_k \to k$; making use of 
\eqs\nr{meas2}, \nr{meas25}, we get  
\be
 \pi \int \! \frac{{\rm d}^d\vec{k}}{(2\pi)^d} 
 \frac{k \delta(\omega - k)}{2} 
 = \frac{\omega^3 \mu^{-2\epsilon}}{4\pi}
 \biggl[
  1 +  \epsilon \biggl(\ln\frac{\bmu^2}{4\omega^2} + 2 \biggr)  
 \biggr]
 \;. 
 \la{fz_basic}
\ee
As far as the $\vec{q}$-integral is concerned, we recall that 
one of the basic 1-loop integrals can, after integration
over $q_0$, be written as
\ba
 B_0(K^2;0,0) & \equiv & 
 \int\!\frac{{\rm d}^D\! Q}{(2\pi)^D}
 \frac{1}{Q^2(K-Q)^2} \la{B0} \\ 
 & = &  
 \int\! \frac{{\rm d}^d\vec{q}}{(2\pi)^d} \frac{1}{4 E_q E_{kq}}
 \biggl[\frac{1}{ik_0 + E_q + E_{kq}} + \frac{1}{-ik_0 + E_q + E_{kq}}\biggr]
 \;. 
\ea
Therefore the 2nd row of \eq\nr{I2_fz0_1} equals 
\be
 \re B_0(-\omega^2+k^2;0,0)
 = \re B_0(-\lambda^2;0,0)
 = \re \biggl[ \frac{\mu^{-2\epsilon}}{(4\pi)^2}
 \biggl( 
   \frac{1}{\epsilon} + \ln\frac{\bmu^2}{-\lambda^2} + 2 
 \biggr) \biggr]
 \;, 
\ee
where we made use of $\delta(\omega-E_k)$ 
in \eq\nr{I2_fz0_1} in order to write
$\omega^2 = k^2 + \lambda^2$. In total, then,  
\ba
 \delta_\rmii{fz,0} \tilde{\mathcal{I}}_2(\omega) & = & 
 \lim_{\lambda\to 0}
 \frac{\omega^3\mu^{-4\epsilon}}{(4\pi)^3}
 \biggl(
   \frac{1}{\epsilon} + \ln\frac{\bmu^2}{4\omega^2}
   + \ln\frac{\bmu^2}{\lambda^2} + 4 \biggr)
 \;. \la{I2_fz0_2}
\ea
Here we observe again the significance 
of the intermediate regulator.

Summing together \eqs\nr{I2_ps0_2}, \nr{I2_fz0_2}, logarithms of 
$\lambda$ cancel against each other, and we obtain the final vacuum 
part of $\tilde{\mathcal{I}}_2$,
given on the first row of \eq\nr{finalnewI2}.

As far as the thermal part is concerned, we change integration variables
on the last row of \eq\nr{I2_1} so that the argument of $\nB{}$ is 
always $E_q$. Furthermore, since the integral is exponentially
convergent, we can use the measure in \eq\nr{meas3}. Partial fractioning
the integrand and inserting $E_k\to\omega$, $E_q\to q$, we get
\ba
 \delta_\rmii{fz,T} \tilde{\mathcal{I}}_2(\omega) & \equiv & 
 \lim_{\lambda\to 0}
 \frac{4\pi}{(2\pi)^3} \int_0^\infty \! {\rm d}k \,  
 \frac{\pi k^3 \delta(\omega - E_k)}{4 E_k} 
 \frac{2\pi}{(2\pi)^3} \int_0^\infty \! {\rm d}q \, {\nB{}(q)}
 \nn & \times & 
 \int_{|k-q|}^{k+q} \! {\rm d}E_{kq} \,
 \mathbbm{P} \biggl[
   \frac{1}{\omega - q + E_{kq}} 
  - \frac{1}{\omega - q - E_{kq}} 
  + \frac{1}{\omega + q + E_{kq}} 
  - \frac{1}{\omega + q - E_{kq}} 
 \biggr]
 \;. \nn \la{I2_fzT_1}
\ea
The integral over $E_{kq}$ is elementary, and sending 
$\lambda\to 0$, such that $\omega =\sqrt{k^2+\lambda^2} \to k$, it yields
\be
 \lim_{\lambda\to 0} \ln\left| 
 \frac{[(\omega-q)^2-(k+q)^2][(\omega+q)^2-(k+q)^2]}
      {[(\omega-q)^2-(k-q)^2][(\omega+q)^2-(k-q)^2]} \right|
 = \ln\left| \frac{q+\omega}{q-\omega} \right|
 \;. 
\ee
In total, then, 
\be
 \delta_\rmii{fz,T} \tilde{\mathcal{I}}_2(\omega) = 
 \frac{\omega^2}{32\pi^3}
 \int_0^\infty \! {\rm d}q \, \nB{}(q) 
 \ln\left| \frac{q+\omega}{q-\omega} \right|
 \;. \la{I2_fzT_2}
\ee
Summing together \eqs\nr{I2_psT_2}, \nr{I2_fzT_2}, we obtain the 
thermal part (second row) of \eq\nr{finalnewI2}.

%
\subsection{An alternative derivation of the vacuum part}
\la{ss:alt}

While the method of sections \ref{ss:mat}--\ref{ss:fz}
works in all cases, the temperature-independent  ``vacuum parts''
of the results can also be derived in a somewhat more 
straightforward manner. We illustrate this 
for the most complicated master integral appearing 
in our study,  $\mathcal{I}_5(\tau)$, defined in \eq\nr{I5_1}.
The topologies from which this structure originates, 
already sketched in graph (j) of \fig\ref{fig:Poly}, 
can be depicted somewhat more specifically as \vspace*{-2mm}
\ba
&& 
 \hspace*{-2.0cm}
 \EleEa \qquad \qquad 
 \EleEb 
 \la{I5_2} \\[1mm] \nonumber 
\ea

The idea of the approach is the following. Going to momentum space; 
carrying out several changes of integration variables; and ignoring possible
problems with Matsubara zero modes, $q_n = 0$ (this will be justified
presently), the integrals over $x$ and $\tau'$ can be carried out, and
we end up with  
\ba
 \mathcal{I}_5(\tau) & = &
 2 \Tint{K} \frac{e^{i k_n\tau}}{K^2}
 \Tint{Q} \frac{1}{Q^2(K-Q)^2} \biggl\{ 
 (2D-5)\vec{k}^2 + (-3D+5) k_n^2 
 \nn & & \; 
 + \frac{2k_n}{q_n} \Bigl[ Q^2 + (D-2) (k_n^2 - \vec{k}^2)\Bigr]
 \biggr\} 
 \;. \la{I5_3}
\ea
We now replace the sum-integral $\Tinti{Q}$ through its zero-temperature
limit, the $D$-dimensional integral $\int_Q$. In addition, the 
Fourier transform over $\tau$ and the subsequent analytic continuation
and cut 
amount, due to the symmetry of the expression in $k_n\to -k_n$, to simply 
setting $k_n \to -i (\omega + i 0^+)$ and taking the imaginary part. 
Thereby any dependence on the temperature disappears. Before the 
analytic continuation, it is useful to write $k_n^2 = K^2 - \vec{k}^2$.

The only complication is the handling of the second row
of \eq\nr{I5_3}. Amusingly, the structure is similar to that met in 
connection with the so-called cyclic Wilson loop in ref.~\cite{pol}.\footnote{%
 In that case the analysis was originally carried out by M.~Veps\"al\"ainen.
 } 
We note, first of all, that in the zero-temperature
limit the Matsubara zero mode plays no special role, and the apparent
divergence $1/q_n$ can be regulated by taking a principal value. 
Furthermore, introducing a Feynman parameter, the integrand can be written as 
\ba
 \frac{2k_n}{q_n} \frac{1}{(k_n-q_n)^2 + (\vec{k-q})^2} & \to &
 \frac{k_n}{q_n}\biggl[ 
 \frac{1}{(k_n-q_n)^2 + (\vec{k-q})^2} - 
 \frac{1}{(k_n+q_n)^2 + (\vec{k-q})^2}
 \biggr] \nn 
 & = & 
 \frac{4 k_n^2}
 {[(k_n-q_n)^2 + (\vec{k-q})^2][(k_n+q_n)^2 + (\vec{k-q})^2]} \nn 
 & = & 
 \int_0^1 \! {\rm d}s \, 
 \frac{4 k_n^2}
 {\bigl\{ (\vec{k-q})^2 + [(2s-1)k_n - q_n]^2 + 4s(1-s) k_n^2  
  \bigr\}^2}
 \;. \nn 
\la{I5_4}
\ea
The discontinuity corresponding
to \eq\nr{I5_3} can now be expressed as 
\ba
 \tilde{\mathcal{I}}_5(\omega) & = &
 2 \im \int_{\vec{k}} \frac{1}{K^2} 
 \Biggl\{ 
   \int_Q \frac{1}{Q^2(K-Q)^2}
   \Bigl[ 5(D-2) \vec{k}^2 + (-3D+5) K^2 \Bigr] 
 \nn & & \; + 
 4 k_n^2 \int_0^1 \! {\rm d}s \, \int_Q \frac{1}{(Q^2 + \tilde M^2)^2}
  \nn & & \; + 
 4 (D-2) k_n^2 (k_n^2 - \vec{k}^2)
 \int_0^1 \! {\rm d}s \, 
 \int_Q \frac{1}{Q^2 \bigl[(\tilde K - Q)^2 + \tilde M^2\bigr]^2}  
 \Biggr\}_{ik_n \to \omega+i0^+}
 \;, \hspace*{0.5cm} \la{I5_5}
\ea
where $\tilde M^2 \equiv 4 s(1-s) k_n^2$ and 
$\tilde K^2 \equiv (2s-1)^2 k_n^2 + \vec{k}^2$.

The $Q$-integrals in \eq\nr{I5_5} are all familiar:
\ba
 \int_Q \frac{1}{Q^2(K-Q)^2} & = & \frac{\mu^{-2\epsilon}}{(4\pi)^2}
 \biggl( \frac{1}{\epsilon} + \ln\frac{\bmu^2}{K^2} + 2 \biggr)
 \;, \\ 
 \int_Q \frac{1}{(Q^2 + \tilde M^2)^2} & = & 
 \frac{\mu^{-2\epsilon}}{(4\pi)^2}
 \biggl( \frac{1}{\epsilon} + \ln\frac{\bmu^2}{\tilde M^2} \biggr)
 \;, \la{Qint2} \\ 
 \int_Q \frac{1}{Q^2 \bigl[(\tilde K - Q)^2 + \tilde M^2 \bigr]^2}  
 & = & 
 \frac{\mu^{-2\epsilon}}{(4\pi)^2\tilde K^2} 
 \ln \frac{\tilde M^2 + \tilde K^2}{\tilde M^2}
 \;. \la{Qint3}
\ea
In the middle case, the remaining integration over the Feynman 
parameter is also straightforward: 
\be
 \int_0^1 \! {\rm d}s \, \ln \frac{\bmu^2}{4 s(1-s) k_n^2}
 = \ln \frac{\bmu^2}{4 k_n^2} + 2 
 \;. \la{Fint1}
\ee

It remains to take the discontinuity and to carry out the integral 
over $\vec{k}$ as well as, where present, over $s$. 
For the structures on the first row of \eq\nr{I5_5} this 
leads directly to the ``known'' cases in \eqs\nr{finalnewI1}, 
\nr{finalnewI2}. We discuss explicitly only the 2nd and 3rd rows 
of \eq\nr{I5_5}, since these structures do not appear elsewhere.

Like in sections~\ref{ss:ps}, \ref{ss:fz}, poles may appear in 
individual terms but are cancelled at the end; to handle these, 
we regulate the expression by replacing $K^2 \to K^2 + \lambda^2$ and
denoting $E_k^2 \equiv k^2 + \lambda^2$. 
Inserting \eqs\nr{Qint2}, \nr{Fint1}, 
the structure on the 2nd row of \eq\nr{I5_5} then becomes
(omitting numerical coefficients)
\ba
 \delta_2 \tilde{\mathcal{I}}_5(\omega) \equiv 
 \frac{\mu^{-2\epsilon}}{(4\pi)^2}
 \int_{\vec{k}}  \im \biggl\{ 
 \frac{k_n^2}{k_n^2+E_k^2}
 \biggl( \frac{1}{\epsilon} + \ln \frac{\bmu^2}{4 k_n^2} + 2 \biggr)
 \biggr\}_{ik_n \to \omega+i0^+}
 \;. 
\ea
We note that, for $\omega > 0$,  
\ba
 \frac{\omega^2}{(-\omega + E_k - i 0^+)(\omega+E_k)}
 & = & \mathbbm{P} \biggl( \frac{\omega^2}{E_k^2 - \omega^2} \biggr)
 + \frac{i \pi \omega}{2} \delta(\omega - E_k)
 \;, \la{4p1} \\ 
 \frac{1}{\epsilon}+\ln\frac{\bmu^2}{4(\omega+i0^+)(-\omega -i0^+)}+2 & = & 
 \frac{1}{\epsilon}+\ln\frac{\bmu^2}{4\omega^2}+2 + i \pi 
 \;. \la{4p2}
\ea
A non-zero contribution only arises by combining the imaginary 
part of \eq\nr{4p1} with the real part of \eq\nr{4p2}, because
the other possibility leads to 
\ba
 \int \! \frac{{\rm d}^d\vec{k}}{(2\pi)^d} 
 \mathbbm{P}\biggl( \frac{1}{k^2 + \lambda^2 - \omega^2}\biggr)
 & = & 
 \int \! \frac{{\rm d}^d\vec{k}}{(2\pi)^d} \frac{1}{k^2}
 \mathbbm{P}\biggl( \frac{k^2 + \lambda^2 - \omega^2 + \omega^2 - \lambda^2}
 {k^2 + \lambda^2 - \omega^2}\biggr)
 \nn & = & 
 \frac{\sqrt{\omega^2-\lambda^2}}{4\pi^2} 
 \int_0^\infty \! {\rm d}k \, \mathbbm{P} \biggl( 
   \frac{1}{k - \sqrt{\omega^2 - \lambda^2}}  - 
   \frac{1}{k + \sqrt{\omega^2 - \lambda^2}}
 \biggr) 
 \nn & = & 0
 \;. \la{Pzero}
\ea
Making use of \eq\nr{fz_basic} in order to evaluate the 
$\vec{k}$-integral then leads to 
\be
 \delta_2 \tilde{\mathcal{I}}_5(\omega) = 
 - \frac{\omega^3 \mu^{-4\epsilon}}{(4\pi)^3}
 \biggl[ 
  \frac{1}{\epsilon} + 2 \biggl(
    \ln\frac{\bmu^2}{4\omega^2} + 2
  \biggr) 
 \biggr] 
 \;. \la{del2I5}
\ee

It remains to consider
the structure on the 3rd row of \eq\nr{I5_5}.
Inserting \eq\nr{Qint3} and shifting the Feynman parameter by $\fr12$
in order to make the integrand symmetric, 
it can be expressed as 
\ba
 \delta_3 \tilde{\mathcal{I}}_5(\omega) \!\! & \equiv & \!\!  
 \frac{2\mu^{-2\epsilon}}{(4\pi)^2}
 \int_0^{\fr12} \!\!\! {\rm d}s \! \int_{\vec{k}} 
 \im \biggl\{ 
  \frac{k_n^2}{k_n^2+E_k^2} \frac{k_n^2 - {k}^2}{(2k_n s)^2 + k^2}
 \biggl( \ln \frac{k_n^2+k^2}{k_n^2} - \ln(1-4s^2) \biggr)
 \biggr\}_{ik_n \to \omega+i0^+}
 \hspace*{-1.6cm} \;. \hspace*{1cm} 
\ea
The integrand is the product of three terms; the first one is 
like in \eq\nr{4p1}, while the 2nd and 3rd terms can for 
$\omega > 0$  be written as 
\be
 \frac{\omega^2+k^2}
 {(-2\omega s + k - i 0^+)(2\omega s+k)}
 =  \mathbbm{P} \biggl( \frac{\omega^2+k^2}{k^2 - 4\omega^2s^2} \biggr)
 + i \pi \theta(\omega-k)  \delta(2\omega s - k) \frac{ \omega^2 +k^2}{2k}
 \;, \la{5p2} 
\ee
\be
 \ln\frac{(-\omega+k-i0^+)(\omega+k)}{(\omega+i0^+)(-\omega -i0^+)}
 -\ln(1-4s^2) =  
 \ln\frac{|k^2-\omega^2|}{\omega^2} -\ln(1-4s^2) + i \pi \theta(k-\omega) 
 \;. \la{5p3}
\ee

Various possible contributions can now be identified, according to
whether the imaginary part comes from  
\eq\nr{4p1}, \nr{5p2}, or \nr{5p3} 
(there is no contribution from multiplying the three
imaginary parts, because of the appearances of $\theta(\omega-k)$
and $\theta(k-\omega)$ in \eqs\nr{5p2}, \nr{5p3}, 
respectively). Actually, the 
imaginary part of \eq\nr{5p2} gives no contribution, because 
the real part of \eq\nr{5p3} vanishes for the value of $s$ that
is allowed by the $\delta$-function of \eq\nr{5p2}. This then 
leaves two possibilities. 

If we take the imaginary part from \eq\nr{5p3} and write 
$\theta(k-\omega) = 1 - \theta(\omega-k)$, then the integral 
over unity can be reduced to integrals like that in \eq\nr{Pzero}
and gives no contribution. A non-zero contribution comes from 
the integral (since the integral is UV-finite we set $d\to 3$)
\ba
 \delta_3^\rmii{\nr{5p3}} \tilde{\mathcal{I}}_5(\omega) 
 \!\! & \equiv & \!\!  
 - \frac{2\pi}{(4\pi)^2}
 \int_0^{\fr12} \!\!\! {\rm d}s \! \int_0^{\omega} \! 
 \frac{{\rm d}k\, k^2}{2 \pi^2}
 \mathbbm{P} \biggl( \frac{\omega^2}{E_k^2 - \omega^2} \biggr)
 \mathbbm{P} \biggl( \frac{\omega^2+k^2}{k^2 - 4\omega^2s^2} \biggr) 
 \nn & = &  
 \frac{\omega^3}{128\pi^3} \int_0^1 \! {\rm d}x \, (1+x^2) 
 \biggl[ 
  \frac{1}{x+1} + \mathbbm{P} \biggl( 
    \frac{1}{x-\sqrt{1-\lambda^2/\omega^2}} \biggr) 
 \biggr]
 \ln\frac{1-x}{1+x} 
 \;, \hspace*{0.7cm} \la{d3aI5}
\ea 
where we carried out the integration over $s$ and denoted $x\equiv k/\omega$; 
the two terms inside the square brackets come from the partial 
fractions of the first principal value (we have set $\lambda\to 0$
wherever the limit exists). 

If we take the imaginary part from \eq\nr{4p1}, the integral becomes
\be
 \delta_3^\rmii{\nr{4p1}} \tilde{\mathcal{I}}_5(\omega) \equiv 
 \frac{\pi\omega}{(4\pi)^2}
 \int_0^{\fr12} \!\!\! {\rm d}s \! \int_0^{\infty} \! 
 \frac{{\rm d}k\, k^2}{2 \pi^2} \delta(\omega - E_k)
 \mathbbm{P} \biggl( \frac{\omega^2+k^2}{k^2 - 4\omega^2s^2} \biggr) 
 \biggl[ 
   \ln\frac{|k^2-\omega^2|}{\omega^2} -\ln(1-4s^2)
 \biggr]
 \;. 
\ee
This time the integral over $k$ is trivial. Denoting $x=2s$, 
a part of the $x$-integral can also be carried out, and we arrive at
\be
 \delta_3^\rmii{\nr{4p1}} \tilde{\mathcal{I}}_5(\omega) = 
 \frac{\omega^3}{64\pi^3}\biggl\{ 
 -\ln\frac{\lambda^2}{4\omega^2} \, \ln\frac{\lambda^2}{\omega^2}
 - \int_0^1 \! {\rm d}x 
 \biggl[ 
  \frac{1}{1+x} + \mathbbm{P} \biggl( 
    \frac{1}{\sqrt{1-\lambda^2/\omega^2}-x} \biggr)    
 \biggr] \ln(1-x^2) 
 \biggr\} 
 \;. \la{d3bI5}
\ee

Adding up \eqs\nr{d3aI5}, \nr{d3bI5}, the coefficient of 
$\ln(1+x)$ develops a prefactor $x^2-1$; this cancels the 
pole (for $\lambda/\omega\to 0$) and 
leaves a regular finite integral. The coefficient 
of $\ln(1-x)$, in contrast, has a prefactor 
$x^2 + 3 = 4 + 2(x-1) + (x-1)^2$, so a singular
integral remains. Denoting $y\equiv \sqrt{1-\lambda^2/\omega^2} - x$ and 
$\delta \equiv 1 - \sqrt{1-\lambda^2/\omega^2} \approx \lambda^2/2\omega^2$, 
the singular integral can be written as 
\ba
 \delta_3^\rmi{s} \tilde{\mathcal{I}}_5(\omega) & \equiv &  
 \frac{\omega^3}{128\pi^3}\biggl\{
  - 4  \int_0^1 \! {\rm d}x  \, 
    \mathbbm{P} \biggl( 
    \frac{1}{\sqrt{1-\lambda^2/\omega^2}-x} \biggr)    
    \ln(1-x)
 \biggr\} 
 \nn & = &
 \frac{\omega^3}{128\pi^3}\biggl\{
  - 4  \biggl[ \int_{-\delta}^{+\delta} \! {\rm d}y  
  + \int_{\delta}^{1-\delta} \! {\rm d}y \biggr]
   \mathbbm{P} \biggl( 
    \frac{1}{y} \biggr)    \ln(y + \delta  ) 
 \biggr\}
  \nn & = &
 \frac{\omega^3}{128\pi^3}\biggl\{
  - 4 \int_0^1 \! \frac{{\rm d}x}{x} \ln \frac{1+x}{1-x}
  - 4 \int_{\delta}^1 \! \frac{{\rm d}y}{y}
 \biggl[ \ln \frac{y+\delta}{y}  + \ln y 
 \biggr]  
 \biggr\} 
 \;,  \la{intstep}
\ea
where in the first integral of the last row
we wrote $x=y/\delta$ and 
in the second could set $\delta\to 0$ in the upper limit
of the integration. Of the remaining integrals the singular
one is now trivially integrated, 
\be
 -4 \int_{\delta}^1 \! \frac{{\rm d}y}{y} \ln y 
 = 2 \ln^2(\delta) \approx 2 \ln^2\biggl(\frac{\lambda^2}{2\omega^2} \biggr)
 \;, 
\ee
whereas it requires a few more steps to show that 
\be
 \lim_{\delta\to 0} \biggl[
 - 4 \int_{\delta}^1 \! \frac{{\rm d}y}{y}
  \ln \frac{y+\delta}{y} \biggr] 
 = -4 \int_0^1 \! \frac{{\rm d}x}{x} \ln ({1+x})
 \;. \la{finalstep}
\ee

Summing up, all terms containing logarithms of $\lambda$
nicely cancel against each other, and we obtain 
\be
  \delta_3 \tilde{\mathcal{I}}_5(\omega) = - \frac{\omega^3}{128\pi^3}
  \biggl( 2 + \frac{4\pi^2}{3} \biggr)
 \;. \la{del3I5}
\ee
We note that the peculiar $\pi^2$-terms, which appear nowhere 
else in our computation, can be traced back to the first term on 
the last row of \eq\nr{intstep} as well as the term in \eq\nr{finalstep}:  
\be
 \int_0^1 \! \frac{{\rm d}x}{x} \ln\frac{ (1+x)^2}{1-x} = \frac{\pi^2}{3}
 \;. 
\ee

Inserting finally the vacuum parts of \eqs\nr{finalnewI1}, 
\nr{finalnewI2} as well as the newly determined 
\nr{del2I5}, \nr{del3I5} into \eq\nr{I5_5}, we obtain 
\be
 \tilde{\mathcal{I}}_5(\omega) = \frac{\omega^3 \mu^{-4\epsilon}}{16\pi^3}
 \biggl(
  \frac{3}{\epsilon} + 6 \ln\frac{\bmu^2}{4\omega^2} + 14 - \frac{8\pi^2}{3}
 \biggr)
 \;. 
\ee
This agrees with the vacuum part of \eq\nr{finalI5} which was independently 
determined with the method of sections \ref{ss:mat}--\ref{ss:fz}.

%
\subsection{Results for all independent sum-integrals}
\la{ss:res}

We list here the results, after Fourier transform (\eq\nr{tildeGE})
and cut (\eq\nr{disc}), for the ``master'' structures of 
\eqs\nr{I5_def}, \nr{I1_def}--\nr{I4_def}. 
The bosonic and fermionic cases differ from each other simply
by the exchange $\nB{} \leftrightarrow - \nF{}$; therefore, we 
only list the bosonic expressions here. The results read:

\ba
 \tilde{\mathcal{I}}_1(\omega) & = & 
 \frac{\omega^3\mu^{-4\epsilon}}{16\pi^3}
 \biggl( \frac{1}{6}
 \biggr)  
 + \frac{1}{16\pi^3} \int_0^\infty \! {\rm d} q \, \nB{}(q) 
 \biggl[
   4 q \omega 
 \biggr] \;, 
\la{finalnewI1} 
\ea

\ba
 \tilde{\mathcal{I}}_2(\omega) & = & 
 \frac{\omega^3\mu^{-4\epsilon}}{16\pi^3}
 \biggl( \frac{1}{4\epsilon} + \fr12 \ln\frac{\bmu^2}{4\omega^2} + \fr{5}3 
 \biggr)  
 \nn & & \; 
 + \frac{1}{16\pi^3} \int_0^\infty \! {\rm d} q \, \nB{}(q) 
 \biggl[
   4 q \omega + \omega^2 \ln \left| \frac{q+w}{q-w} \right| 
 \biggr] \;, 
 \la{finalnewI2} 
\ea

\ba
 \tilde{\mathcal{I}}_3(\omega) & = & 
 \frac{\omega^3\mu^{-4\epsilon}}{16\pi^3}
 \biggl( \frac{1}{24\epsilon} + \fr1{12} \ln\frac{\bmu^2}{4\omega^2} + \fr29 
 \biggr)  
 \nn & & \; 
 - \frac{1}{16\pi^3} \int_0^\infty \! {\rm d} q \, \nB{}(q) 
 \biggl[
   q^2 \ln \left| \frac{q+w}{q-w} \right| 
   + q \omega \ln \frac{|q^2 - \omega^2|}{\omega^2}
 \biggr] \;, 
 \la{finalI3} 
\ea

\ba
 \tilde{\mathcal{I}}_4(\omega) & = & 
 \frac{\omega^3\mu^{-4\epsilon}}{16 \pi^3}
 \biggl( \frac{1}{2\epsilon} + \ln\frac{\bmu^2}{4\omega^2} + \fr{23}{6} 
 \biggr)  
 \nn & & \; 
 - \frac{1}{16\pi^3} \int_0^\infty \! {\rm d} q \, \nB{}(q) 
 \biggl[
   4 q \omega + 
  \mathbbm{P}\biggl( \frac{2q \omega^3 }{\omega^2 - q^2} \biggr)
 \biggr] \;, 
 \la{finalI4} 
\ea

\ba
 \tilde{\mathcal{I}}_5(\omega) & = & 
\frac{\omega^3 \mu^{-4\epsilon}}{16\pi^3}
 \biggl(
  \frac{3}{\epsilon} + 6 \ln\frac{\bmu^2}{4\omega^2} + 14 - \frac{8\pi^2}{3}
 \biggr)
 \nn & & \; 
 + \frac{1}{16\pi^3} \int_0^\infty \! {\rm d} q \, \nB{}(q) 
 \biggl\{
   24 q \omega + 
   \mathbbm{P}\biggl( \frac{8q \omega^3 }{\omega^2 - q^2} \biggr)
   + 20 \omega^2 \ln \left| \frac{q+w}{q-w} \right| 
  \nn & & \hspace*{3cm} +\; \frac{8 \omega^4}{q} 
     \biggl[\frac{1}{q+\omega} \ln \frac{q + \omega}{\omega} 
     + \mathbbm{P} \biggl(  \frac{1}{q-\omega} \ln \frac{\omega}{|q - \omega|} \biggr) 
     \biggr]
 \biggr\} \;. \hspace*{0.7cm}
 \la{finalI5}
\ea
Note that the integral 
$
 \int_0^\infty \! {\rm d}q \, q \, \nB{}(q)
$
appearing frequently could be carried out 
analytically (cf.\ \eq\nr{nB_asy_int}), but many others
cannot so we choose to display all thermal contributions
in a uniform unintegrated form.

%
\section{Details of Hard Thermal Loop resummation}

We present here the details of the Hard Thermal Loop resummation, 
outlined and motivated in \se\ref{ss:HTL}. Concretely, 
the computations below amount to a derivation of \eq\nr{rho_soft}.

\subsection{Framework} 

At $\omega\ll T$ the unresummed result 
of \eq\nr{rho_hard} is logarithmically infrared divergent
if we try to take the intercept determining 
the momentum diffusion coefficient $\kappa$ defined 
in \eq\nr{kappa_def_3} 
(cf.\ \eq\nr{kappa_hard}). In order to obtain a result that
is valid in this regime as well, the perturbative series 
needs to be reorganized. The well-known framework for achieving this
is that of the Hard Thermal Loop (HTL) 
effective theory~\cite{htl}. In order to avoid double-counting, 
we first need to subtract the unresummed version of the HTL
contribution from the QCD result, whereby we should obtain 
a result that is infrared finite but not physical, since 
the contribution of the soft modes, $k\sim gT$, is not included
(cf.\ \eq\nr{resum}); 
subsequently the resummed HTL contribution is added, whereby the result 
remains infrared finite and now involves the physical contribution 
of the Debye scale as well. For both purposes, we start
by recalling the HTL form of the gluon self-energy in $D$ dimensions 
(at $\rmO(g^4)$ only the self-energy contribution is needed
within the HTL theory, cf.\ section~3 of ref.~\cite{eucl}).

Carrying out the Matsubara sums over $q_n$ and $\{ q_n \}$
in the self-energy of \eq\nr{Pi_xi}; 
expanding the result to leading order
in $ik_n/T$, $k/T$;  and restricting for the moment to spatial 
components, one finds
\be
 \Pi_{ij}(K) = g^2 (D-2) 
 \int_{\vec{q}} \frac{1}{q} 
 \Bigl[ 2 \Nf\, \nF{}(q) + (D-2) \Nc\, \nB{}(q) \Bigr] 
 \frac{v_i v_j ik_n}{ik_n - \vec{k}\cdot\vec{v}}
 + g^2 \rmO\biggl(\frac{ik_n,k}{T} \biggr)
 \;, 
\ee
where $\vec{v} \equiv \vec{q}/q$. Introducing 
the independent projection operators 
\be
 P^T_{\mu\nu}(K) \equiv \delta_{\mu i} \delta_{\nu j}
 \biggl( \delta_{ij} - \frac{k_i k_j}{k^2} \biggr)
 \;, \quad
 P^E_{\mu\nu}(K) \equiv 
 \delta_{\mu\nu} - \frac{K_\mu K_\nu }{K^2} - P^T_{\mu\nu}(K)
 \;, 
\ee
and writing $\Pi_{ij} = P^T_{ij} \, \Pi_T + P^E_{ij} \,  \Pi_E$, 
the scalar structures $\Pi_T$, $\Pi_E$ can be projected out: 
\ba
 \Pi_E(K) & = & 
  g^2(D-2)\frac{K^2}{k^2} \int_{\vec{q}} 
  \! \frac{2 \Nf \, \nF{}(q) + (D-2) \Nc \, \nB{}(q)}{q}
  \frac{\vec{k}\cdot\vec{v}}{ik_n + \vec{k}\cdot\vec{v}}
  \;, \la{PiE} \\ 
  \Pi_T(K) & = &
  g^2 \int_{\vec{q}} 
  \! \frac{2 \Nf \, \nF{}(q) + (D-2) \Nc \, \nB{}(q)}{q}
  \biggl( 1 - \frac{K^2}{k^2} 
  \frac{\vec{k}\cdot\vec{v}}{ik_n + \vec{k}\cdot\vec{v}}
  \biggr) 
  \;. \la{PiT} 
\ea
For future reference, we also introduce the notation 
\be
 \mE^2 \equiv g^2 (D-2) \int_{\vec{q}} \frac{1}{q} 
 \Bigl[ 2 \Nf \nF{}(q) + (D-2) \Nc \nB{}(q) \Bigr]
 \;, \la{mmE}
\ee
which for $D=4$ reproduces the known expression for 
the Debye mass parameter squared, given in \eq\nr{mE}.

\subsection{Naive HTL analysis}

We first carry out an {\em unresummed} HTL computation, which 
needs to be subtracted from the naive QCD result in order to avoid
double-counting and in order to obtain an infrared-finite result for 
the ``matching coefficient'' part of the final expression 
(cf.\ \eq\nr{resum}). 
This means that the self-energy is treated as an ``insertion'', 
i.e.\ as if it were an interaction. We also note that since the HTL 
theory only represents the infrared physics of QCD and has not
been derived with relative accuracy $\rmO(g^2)$, the gauge coupling
appearing in it is to be treated as the renormalized coupling. 
Then the unresummed result contains a free contribution like 
in \eq\nr{rhoE2}, 
\be
 \Bigl[ \rho_\rmii{E}^{(2)}(\omega) \Bigr]_\rmii{HTL, naive} 
 = \frac{g^2 C_F}{6\pi} \omega^3
 \;, \la{rho2_HTL_unres}
\ee
as well as a next-to-leading order term which we first write
in configuration space, 
\be 
 \Bigl[ G_\rmii{E}^{(4)}(\tau) \Bigr]_\rmii{HTL, naive} 
 = 
 \frac{g^2C_F}{3}  \Tint{K} 
 \frac{e^{i k_n\tau}}{(K^2)^2} 
 \Bigl[
   (D-2) \, k_n^2 \, \Pi_T(K) + K^2 \, \Pi_E(K) 
 \Bigr] 
 \;. \la{GE4_HTL_unres}
\ee
To arrive at this expression we started from the 
form in \eq\nr{selfE} and inserted 
$
 \Pi_{\mu\nu} = P^T_{\mu\nu} \, \Pi_T + P^E_{\mu\nu} \,  \Pi_E
$. 
Plugging in $\Pi_T$, $\Pi_E$ from \eqs\nr{PiE}, \nr{PiT} 
and partial fractioning the dependence on $ik_n$, we get 
\ba
 && \hspace*{-1cm} 
 \Bigl[ G_\rmii{E}^{(4)}(\tau) \Bigr]_\rmii{HTL, naive} 
 = \frac{g^4 C_F (D-2)}{3}
   \int_{\vec{q}} \frac{2 \Nf \, \nF{}(q) + (D-2) \Nc \, \nB{}(q)}{q}
 \Tint{K} e^{i k_n\tau}
 \biggl[
  \frac{2}{K^2} - \frac{k^2}{(K^2)^2}
 \nn & & \; 
  + \, \frac{1}{2(k-\vec{k}\cdot\vec{v})}
    \biggl( \frac{1}{ik_n + \vec{k}\cdot\vec{v}} - 
            \frac{1}{ik_n + k} \biggr) 
  - \frac{1}{2(k+\vec{k}\cdot\vec{v})}
    \biggl( \frac{1}{ik_n + \vec{k}\cdot\vec{v}} - 
            \frac{1}{ik_n - k} \biggr) 
 \biggr]
 \;. \nn \la{GE4_HTL_1}
\ea
It remains to Fourier transform (cf.\ \eq\nr{tildeGE})
and take the discontinuity (cf.\ \eq\nr{disc}). 
For the structures at the end of the first
row of \eq\nr{GE4_HTL_1} this goes like before 
(cf.\ \eq\nr{fz_basic}) and yields, for $D\to 4$, 
\be
 \Tint{K} \frac{e^{i k_n\tau}}{K^2} \to \frac{\omega}{4\pi}
 \;, \quad
 \Tint{K} \frac{k^2e^{i k_n\tau}}{(K^2)^2} \to \frac{3\omega}{8\pi}
 \;, \quad
 \Tint{K} {e^{i k_n\tau}}
 \biggl[
  \frac{2}{K^2} - \frac{k^2}{(K^2)^2}
 \biggr]
 \to \frac{\omega}{8\pi}
 \;. \la{simple}
\ee
For the terms on the second row of \eq\nr{GE4_HTL_1}, 
on the other hand, the corresponding steps amount to 
\be
 T\sum_{k_n} \frac{e^{i k_n\tau}}{ik_n + \Delta}
 \to \pi \delta(\omega - \Delta)
 \;, 
\ee
but the remaining integration over the spatial components of $\vec{k}$ 
requires regularization, so we have to proceed carefully.  

Making use of the integration measure in \eq\nr{meas2}, save for 
$\vec{k}$, with the polar axis chosen in the direction of 
$\vec{v} = \vec{q}/q$
and denoting $z\equiv \vec{k}\cdot\vec{v}/k$, 
we are lead to consider the integral 
($d\equiv D-1$)
\be
 \mathcal{I}_\rmii{HTL}(\omega) \equiv
 \int_0^{\infty} \! {\rm d}k 
 \int_{-1}^{+1} \! {\rm d}z \, 
 \frac{ k^{d-2} (1-z^2)^{\frac{d-3}{2}}}
      {2(4\pi)^{\frac{d-1}{2}} \Gamma(\frac{d-1}{2})}
 \biggl\{ 
   \frac{1}{1-z}\Bigl[ \delta(\omega - kz) -\delta(\omega - k) \Bigr]
  - 
   \frac{1}{1+z}\Bigl[ \delta(\omega - kz) \Bigr]
 \biggr\} 
 \;, 
\ee
where a term with $\delta(\omega+k)$ was omitted 
because of a restriction to $\omega > 0$. 
In the term with $\delta(\omega - k)$ 
we can trivially integrate over $k$; 
in the terms with $\delta(\omega - kz)$ 
we can trivially integrate over $z$, noting 
that the constraints can get realized only for $k > \omega$. 
Renaming the integration variable subsequently as $k = \omega z$,
$z > 1$,  
and setting $d=3-2\epsilon$, the result can be expressed as 
\be
 \mathcal{I}_\rmii{HTL}(\omega) = 
 \frac{\omega^{1-2\epsilon}}
      {2(4\pi)^{1-\epsilon}\Gamma(1-\epsilon)}
 \biggl[
   \int_1^\infty \!\! {\rm d}z \, (z^2-1)^{-\epsilon}
   \biggl( \frac{1}{z-1} + \frac{1}{z+1} \biggr)
   -  
   \int_{-1}^{+1} \!\! {\rm d}z \, \frac{(1-z^2)^{-\epsilon}}{1-z}
 \biggr]
 \;.
\ee
The first integral here can be rewritten as 
$
 \int_1^\infty \! {\rm d}y (y-1)^{-1-\epsilon}
$ 
with $y=z^2$, and as an analytic function it vanishes. 
The second integral inside the square brackets evaluates to 
$
 1/\epsilon - \ln 4 + \rmO(\epsilon)
$.
Expanding the prefactor as well and introducing the $\msbar$
scale parameter as before, we obtain 
\be
 \mathcal{I}_\rmii{HTL}(\omega) = 
 \frac{\omega\mu^{-2\epsilon}}{8\pi}
 \biggl(
  \frac{1}{\epsilon} + \ln\frac{\bmu^2}{4\omega^2} 
 \biggr) 
 \;. 
\ee
Returning to \eq\nr{GE4_HTL_1} and inserting $\mE^2$ from 
\eq\nr{mmE} as well as the simple structures from \eq\nr{simple}, 
we finally arrive at 
\be
 \Bigl[ \rho_\rmii{E}^{(4)}(\omega) \Bigr]_\rmii{HTL, naive}
 = \frac{g^2 C_F \mE^2}{3}
 \frac{\omega\mu^{-2\epsilon}}{8\pi}
 \biggl(
  \frac{1}{\epsilon} + \ln\frac{\bmu^2}{4\omega^2} + 1
 \biggr)  
 \;. \la{rhoE_HTL_unres}
\ee

\subsection{Resummed intercept}

The remaining step is to carry out the {\em resummed} 
HTL computation (the last term in \eq\nr{resum}). 
As it happens, this problem was already
addressed in ref.~\cite{eucl}, albeit for the difference 
$
 \rho_\rmii{E}(\omega)/\omega - 
 \lim_{\omega\to 0} \rho_\rmii{E}(\omega)/\omega
$; 
the result is reproduced in \eq\nr{wdep}. 
In the following, it is then our task to consider
the intercept 
$
 \lim_{\omega\to 0} \rho_\rmii{E}(\omega)/\omega
$, 
which was not addressed in ref.~\cite{eucl}.
In the following we {\em define} 
$[\rho_\rmii{E}]_\rmii{IR}$ to be the spectral function computed
within the HTL theory, but only up to linear order in a 
Taylor expansion in $\omega$ around zero, 
i.e.\ corresponding to the intercept 
$\lim_{\omega\to 0} \rho_\rmii{E}(\omega)/\omega$.
It turns out that this limit is logarithmically ultraviolet 
divergent within the HTL theory, with the same $1/\epsilon$
as in \eq\nr{rhoE_HTL_unres}. 

The computation proceeds somewhat analogously to the unresummed one 
above, only now \eqs\nr{rho2_HTL_unres}, \nr{GE4_HTL_unres}, as well as 
higher order corrections are resummed. Writing the expression 
after the Fourier transform in \eq\nr{tildeGE}, 
and making use of the evenness of $\Pi_T, \Pi_E$ in 
$k_n\to -k_n$, we get 
\be
 \Bigl[ \tilde G_\rmii{E}^{(4)}(\omega_n)\Bigr]_\rmii{HTL, resum} = 
 - \frac{g^2 C_F}{3}
 \int_{\vec{k}} 
 \biggl[
   \frac{(D-2)\omega_n^2}{\omega_n^2 + {k}^2 + \Pi_T(\omega_n,\vec{k})} +  
   \frac{\omega_n^2 + {k}^2}
    {\omega_n^2 + {k}^2 + \Pi_E(\omega_n,\vec{k})} 
 \biggr] 
 \;. \la{tGE_IR_1}
\ee
Taking now the discontinuity (\eq\nr{disc}) and noting that the 
first term of \eq\nr{tGE_IR_1} involves a high power of $\omega$, 
the contribution to the part linear in 
$\omega$ must come from the second term. 
Recalling \eq\nr{PiE}, we observe that 
$\omega_n^2 + k^2$ nicely cancels out. 
For small $\omega$ we can furthermore write
the structure inside $\Pi_E$ as
\be
 \frac{\vec{k}\cdot\vec{v}}{-\omega + \vec{k}\cdot\vec{v} - i 0^+}
 = 
 1 + \mathbbm{P}\biggl(\frac{\omega}{-\omega + \vec{k}\cdot\vec{v}}\biggr)
 + i \omega\pi\delta(\omega - \vec{k}\cdot\vec{v})
 \approx 
 1 + i \omega\pi\delta(\vec{k}\cdot\vec{v})
 \;, 
\ee
where in the real part the principal value integral leads
to a contribution vanishing faster for $\omega\to 0$ than the unity. 
The unity leads directly to the $\mE^2$ of \eq\nr{mmE} and so we get
\be
 \Bigl[ \rho_\rmii{E}^{(4)}(\omega) \Bigr]_\rmii{IR, resum}  
 = \frac{g^4 C_F (D-2)\omega}{3}
   \int_{\vec{q}} \frac{2 \Nf \, \nF{}(q) + (D-2) \Nc \, \nB{}(q)}{q}
   \int_{\vec{k}} \frac{\pi k^2\delta(\vec{k}\cdot\vec{v})}
  {(k^2 + \mE^2)^2}
 \;. \la{rhoE_IR_2}
\ee
The $\vec{k}$-integral here can again be carried out by choosing 
$\vec{v}$ as the polar axis and making use of the measure
in \eq\nr{meas2}:
\ba
 \mathcal{I}_\rmii{IR} & \equiv & \pi \int \! 
 \frac{{\rm d}^d\vec{k}}{(2\pi)^d}
 \frac{k^2\delta(\vec{k}\cdot\vec{v})}
  {(k^2 + \mE^2)^2}  
 \nn & = & 
 \frac{1}
      {(4\pi)^{\frac{d-1}{2}} \Gamma(\frac{d-1}{2})}
 \int_0^{\infty} \! {\rm d}k \, \frac{k^{d+1}}{(k^2 + \mE^2)^2}
 \int_{-1}^{+1} \! {\rm d}z \, (1-z^2)^{\frac{d-3}{2}} 
 \delta(k\,z)
 \nn & = &  
 \frac{\mE^{-2\epsilon}}{2(4\pi)^{1-\epsilon}\Gamma(1-\epsilon)}
 \int_0^\infty \!\! {\rm d}y \, \frac{y^{1-\epsilon}}{(y+1)^2}
 \;, 
\ea
where $y\equiv k^2/\mE^2$. Expanding in $\epsilon$ we obtain 
\be
 \mathcal{I}_\rmii{IR} = 
 \frac{\mu^{-2\epsilon}}{8\pi}
 \biggl(
  \frac{1}{\epsilon} + \ln\frac{\bmu^2}{\mE^2} - 1 
 \biggr) 
 \;,
\ee
and \eq\nr{rhoE_IR_2} in combination with \eq\nr{mmE} then yields
\be
 \Bigl[ \rho_\rmii{E}^{(4)}(\omega) \Bigr]_\rmii{IR, resum}
 = \frac{g^2 C_F \mE^2}{3}
 \frac{\omega\mu^{-2\epsilon}}{8\pi}
 \biggl(
  \frac{1}{\epsilon} + \ln\frac{\bmu^2}{\mE^2} - 1
 \biggr)  
 \;. \la{rhoE_IR_resum}
\ee

Combining (as dictated by \eq\nr{resum})
the unresummed HTL result 
(\eqs\nr{rho2_HTL_unres}, \nr{rhoE_HTL_unres}) 
with the resummed IR limit (\eq\nr{rhoE_IR_resum}) 
we arrive at \eq\nr{rho_soft}, from 
which $1/\epsilon$-divergences have duly cancelled.

\subsection{Resummed $\omega$-dependence}

The remaining ingredient, 
the HTL-resummed contribution beyond the linear term in $\omega$, i.e.\ 
$
  \bigl[ \rho_\rmii{E}(\omega) \bigr]_\rmii{HTL, resum}  - 
  \bigl[ \rho_\rmii{E}(\omega) \bigr]_\rmii{IR, resum}
$, 
was determined in ref.~\cite{eucl}, and is reproduced 
in \eq\nr{wdep} of the main text.
The four terms correspond to the transverse cut, electric cut, 
transverse pole, and electric pole, respectively. 
The $\hat\omega$-independent subtraction 
in the electric cut corresponds to \eq\nr{rhoE_IR_2}.


\end{document}